\documentclass[12pt,preprint]{aastex}

\shorttitle{GK stars NLTE UV band}
\shortauthors{Short, Campbell, Pickup \& Hauschildt}

\begin{document}


\title{Modeling the near-UV band of GK stars, Paper II: NLTE models}


\author{C. Ian Short}
\affil{Department of Astronomy \& Physics and Institute for Computational Astrophysics, Saint Mary's University,
    Halifax, NS, Canada, B3H 3C3}
\email{ishort@ap.smu.ca}

\author{Eamonn A. Campbell}
\affil{Department of Astronomy \& Physics and Institute for Computational Astrophysics, Saint Mary's University,
    Halifax, NS, Canada, B3H 3C3}
\email{}

\author{Heather Pickup}
\affil{Department of Physics \& Astronomy, University of Waterloo,
    Waterloo, ON, Canada }
\email{}

\author{Peter H. Hauschildt}
\affil{Hamburger Sternwarte, Gojenbergsweg 112, 21029 Hamburg, Germany}
\email{yeti@hs.uni-hamburg.de}




\begin{abstract}

  We present a grid of atmospheric models and synthetic spectral energy distributions (SEDs) for 
late-type dwarfs and giants of solar and 1/3 solar metallicity with many
opacity sources computed in self-consistent Non-Local Thermodynamic Equilibrium (NLTE), and compare them
to the LTE grid of \citet{shorth10} (Paper I).  We describe, for the first time, how the NLTE
treatment affects the thermal equilibrium of the atmospheric structure ($T(\tau)$ relation)
and the SED as a finely sampled function of $T_{\rm eff}$, $\log g$, and $[{{\rm A}\over {\rm H}}]$ among solar 
metallicity and mildly 
metal poor red giants.  We compare the computed SEDs to the library of observed 
spectrophotometry described in Paper I across the entire visible band, and 
in the blue and red regions of the spectrum separately.  
We find that for the giants of both metallicities, the NLTE models yield best fit $T_{\rm eff}$
values that are ~30 to 90 K lower than those provided by LTE models, while providing greater 
consistency between $\log g$ values, and, for Arcturus, $T_{\rm eff}$ values, fitted 
separately to the blue and red spectral regions.  There is marginal evidence that NLTE models
give more consistent best fit $T_{\rm eff}$ values between the red and blue bands for 
earlier spectral classes among the solar metallicity GK giants than they do for the later classes,
but no model fits the blue band spectrum well for any class.  
For the two dwarf spectral classes that we are able to study, the effect of NLTE on 
derived parameters is less significant.  We compare our derived $T_{\rm eff}$ values to
several other spectroscopic and photometric $T_{\rm eff}$ calibrations for red giants, 
including one that is less model dependent based on the infrared flux method (IRFM).  
We find that the NLTE models provide slightly better agreement to the IRFM calibration
 among the warmer stars in our sample, while giving approximately the same level
of agreement for the cooler stars. 

\end{abstract}



\section{Introduction}



  Previously, we have compared the quality
of fit provided by atmospheric models, high resolution synthetic spectra, and spectral energy distributions 
(SEDs, $f_{\lambda}(\lambda)$) computed both in LTE, and with many opacity sources treated in self-consistent 
Non-LTE (NLTE),
for the the Sun and the standard stars Procyon ($\alpha$ UMi) and Arcturus ($\alpha$ Boo) (\citep{shorth09}, 
\citep{shorth05}, \citep{shorth03}).  
We found that our LTE models tend to increasingly predict too much blue and near-UV band flux as
$T_{\rm eff}$ decreases, and that the problem is exacerbated by non-LTE effects (mainly the
non-LTE over-ionization of \ion{Fe}{1}, as is well explained in the case of the Sun (see,
for example, \citet{rutten86})).  However, their conclusions were weak because of the small number of stars
covering a few haphazard points in stellar parameter space ({$T_{\rm eff}/\log g/[{{\rm A}\over{\rm H}}]$}).
\citet{shorth10} (hereafter Paper I) took a first step toward making the investigation more comprehensive 
by comparing a large grid of LTE model 
SEDs spanning the cool side of the HR diagram to observed SEDs taken from the extensive uniformly re-calibrated
spectrophotometric catalog of
\citet{burn}.  They investigated LTE models and synthetic SEDs computed with two choices of input atomic lines list:
a larger, lower quality ``big'' list, and a smaller, higher quality ``small'' list, and found that the models 
computed with the ``small'' line list provide
greater internal self-consistency among different spectral bands, and
closer agreement with the less model-dependent $T_{\rm eff}$ scale of \citet{ramirezm05},
but not to the interferometrically derived $T_{\rm eff}$ values of
\citet{baines10}.  
 They also found that to within the limits of the observed spectrophotometry, there was
no evidence of a systematic over-prediction of blue and near-UV band flux among GK giants 
in general, but they did confirm the over-prediction for Arcturus (their ``K1.5III-0.5'' sample).  

\paragraph{}
Here we take the next step 
by carrying out a similar comparison for a large grid of models SEDs with many important extinction 
sources treated in self-consistent NLTE (see \citet{shorth03} for a description of these 
atmospheric models and spectra with H, He, and two or more of the lowest ionization stages of
C, N, O, and most of the light metals and the Fe-group elements treated in self-consistent multi-species
non-LTE statistical equilibrium.)  Our goal is to map out the goodness of fit, and
the magnitude of any systematic discrepancies between model and observed SEDs, as a
function of the three stellar parameters, $T_{\rm eff}$, $\log g$, and
$[{{\rm A}\over{\rm H}}]$, this time for NLTE models, and to compare the
results to those of LTE modeling.  We also compare
our $T_{\rm eff}$ values inferred from SED fitting to less model-dependent 
$T_{\rm eff}$ calibrations.  One important goal is to determine
where in the upper right quadrant of the HR diagram NLTE effects become most important.


\section{Observed $f_{\lambda}(\lambda)$ distributions \label{sobsseds}}

\citet{burn} presented a large catalog (henceforth B85) of observed SEDs taken with 
photo-electric instruments on 0.5m class telescopes at 
various observatories in the former USSR from the late 1960s to the mid 
1980s, and uniformly 
photometrically re-calibrated to the ``Chilean system''.  \citet{shorth09} 
contains a more detailed description of the individual data sources
included in this compilation.    
These data sets
all generally cover the $\lambda$ range 3200 to 8000 \AA~ with 
a nominal sampling, $\Delta\lambda$, of 25 \AA, and have a quoted ``internal
photometric accuracy'' of $\approx3.5\%$.  
A point worth repeating from Paper I is that to match the appearance of the synthetic
to the observed spectra, we had to convolve the synthetic spectra with an
instrumental broadening kernel corresponding to a resolution element, $\Delta\lambda$, of 75 A. 

\paragraph{}

Paper I contains a description of our procedure for extracting quality-controlled samples 
of spectra from the B85 catalog and forming mean and $\pm 1 \sigma$ deviation spectra
for each spectral type at each $[{{\rm A}\over{\rm H}}]$ value.  We note here for the first
time that our procedure effectively yields a useful spectrophotometric library for solar metallicity 
GK stars.  To briefly summarize, 
the procedure involves cross-referencing the B85 catalog with the 5$^{\rm th}$ Revised Edition of the Bright Star 
Catalog \citep{hoffleitw91}, henceforth BSC5) to screen out stars flagged as exhibiting binarity, 
chemical peculiarity, or
variability of any kind.  The B85 catalog does not contain metallicity information, therefore, we then 
identified our B85 stars in the metallicity catalog of \citet{cayrelsr01}.  For many, but not all, 
of our stars, the \citet{cayrelsr01} contains multiple $[{{\rm A}\over{\rm H}}]$ values.  For objects 
where these were approximately randomly distributed, we found the mean metallicity.  For objects
where these had a skewed distribution, we disregarded the deviant values (usually only one)), and found a 
modal metallicity.  We only retained stars for which the mean (or modal) $[{{\rm A}\over{\rm H}}]$ value was 
within $\pm 0.1$ of either of our two nominal $[{{\rm A}\over{\rm H}}]$ values of
interest (0.0 and -0.5).  

\paragraph{}

Spectral and luminosity classes were finalized by cross-referencing B85 stars with The Revised Catalog 
of MK Spectra Types for the Cooler Stars \citep{keenan_n00}, the paper of \citet{keenan_b99}, 
The Perkins Catalog of Revised MK Types for the Cooler Stars \citet{keenan_m89}, or \citet{skiff10},
 in decreasing order of preference.  We also formed mean $B-V$ values for our spectral types by
cross-referencing B85 stars with the Catalog of Homogeneous Means in the UBV System \citep{mermilliod91}.
(As a result, we found the BSC5 catalog to accurately reflect the primary sources for these stars, and 
could have relied largely on it alone for spectral types and colors.)   
All spectra were
corrected for their heliocentric radial velocity, {\it RV}, using the values in
BSC5.  However, we expect the {\it RV} correction to have a very minor effect on the quality 
of spectral fitting at the low spectral resolution of the B85 data. 

\paragraph{}

In keeping with our automated approach, we make no attempt to 
find values in the literature (of possibly variable quality) for the 
distance and radius of each star. 
Rather, all spectra have been interpolating to a common regular $\lambda$ grid, and then
a ``quasi-bolometric'' normalization was applied by dividing
them by the entire area under the spectrum from 3200 to 7500 \AA.  
We note that this differs from the normalization used in Paper I, in which the 
spectra were forced to have the same flux in a narrow spectral region around 6750 \AA.  
We suspect that the normalization used in Paper I may artificially enhance the quality of fit
at the red end of the spectrum with respect to that at the blue end, and is overly
reliant on the absence of any unexpected features around 6750 \AA. 
For each spectral type and $[{{\rm A}\over{\rm H}}]$ value, we calculate mean and
$\pm 1 \sigma$ deviation spectra for the sample of corresponding individual 
spectra.  Table 1 of Paper I shows how many stars of each spectral/class and 
$[{{\rm A}\over{\rm H}}]$ value, and the number of spectra per star, were
finally retained from the B85 catalog, along with the identities of the stars.  In Table \ref{tabb85} we present 
summary information showing the total number of observed spectra that were used to form the
mean and $\pm 1 \sigma$ deviation spectra in each spectral class/$[{{\rm A}\over{\rm H}}]$ sample.  
Any individual spectra that deviated by more than $\approx 1 \sigma$
from the sample mean over a significant $\lambda$ range were rejected and
the mean and deviation spectra were re-calculated.
This resulted in a final set of 44 spectra of 33 stars, 30 of 
$[{{\rm A}\over{\rm H}}]=0.0$ and three of $[{{\rm A}\over{\rm H}}]=-0.5$.  
Fig. \ref{fvarymult} shows the comparison of the sample mean and $\pm 1 \sigma$ deviation spectra to the
distribution of individual spectra for the illustrative case of the G8 III/$[{{\rm A}\over{\rm H}}]=0.0$ sample.
   
\paragraph{Arcturus. }

We note that our
K1.5 III sample of $[{{\rm A}\over{\rm H}}]=-0.5$ consists entirely of three measurements of
the spectrum of Arcturus.  Therefore, the evaluation of NLTE and LTE fits to this sample is
directly comparable to the NLTE modeling of Arcturus of \citet{shorth03} and \citet{shorth09}.

\section{Model grid}

\subsection{Atmospheric structure calculations}

 The grid of LTE spherical atmospheric models and synthetic SEDs computed with PHOENIX V. 150303C, covering 
$\approx$ 600
parameter points, was described in detail in Paper I.  The most pertinent point
to reiterate here is that the grid has sampling intervals, $\Delta T_{\rm eff}$,
of 125 K and $\Delta \log g$ of 0.5.   
The grid covers $\log g$ values from
3.0 to 1.5 at all $T_{\rm eff}$ values from 4000 to 5625 K, goes to down to 1.0 for all models
of $T_{\rm eff} \le 5000$ K, and includes values from 4.0 to 5.0 for $T_{\rm eff} \ge 5375$ K.  
All models are computed at $[{{\rm A}\over{\rm H}}]$ values of 0.0 and -0.5.
The radii of these spherical
models were determined by holding the mass fixed at $1 M_{\rm Sun}$, and the justification is
described in Paper I and more extensively in the careful investigation of PHOENIX LTE models of red giants
in the ``NextGen'' grid of \citet{hauschildtafba99}.  
The value adopted for the micro-turbulent velocity dispersion, $\xi_{\rm T}$, 
increases from 1 to 4 km s$^{-1}$ as $\log g$ decreases.  Based on numerical experiments 
with $\xi_{\rm T}$ values of 2 and 4 km s$^{-1}$ at $T_{\rm eff}=4000$ K, $\log g=1.0$, and
$[{{\rm A}\over{\rm H}}]=0.0$, we
find that the value has little discernible impact on the synthetic SEDs once they are convolved to
match a spectral resolution element, $\Delta\lambda$, of 75 A. 
The atmospheres of GK stars become convective below a continuum optical depth of unity.  PHOENIX
employs the B\" oehm-Vitense mixing-length theory (MLT) of convection, and we adopted
a mixing length parameter for the treatment of convective flux transport of one pressure
scale height.  Given the scope
of the model grid required for this initial investigation, we have decided to restrict ourselves to
scaled solar $[{{\rm A}\over{\rm H}}]$ distributions, with the solar
abundance distribution of \citet{grev_ns92}.  The considerations leading to this choice were
discussed in Paper I, but are worth reiterating here given the recent discussion surrounding
solar abundances (see, for example, \citet{asplund04}).  There has been some tension between
3D NLTE spectroscopic abundances and helioseismological abundances that makes it difficult to
clarify which abundances to prefer.  We plan to extend our investigation in the future by 
exploring the effects of both alternate solar abundances, and non-solar abundances for metal-poor
stars.    

\paragraph{}

We note again here that our models are in hydrostatic and radiative/convective equilibrium, and are static and
horizontally homogeneous.  Therefore, they cannot account for the effects of chromospheric heating, nor for
star spots, active regions, granulation, or other horizontal inhomogeneities.

\subsubsection{NLTE}

\citet{shorth05} contains a description of the method and scope of the NLTE statistical equilibrium (SE)
treatment in PHOENIX and the sources of critical atomic data, and we only re-iterate the most pertinent aspects 
here.  If necessary, PHOENIX can include at least the lowest two stages of 24 elements,
including the lowest six ionization stages of the 20 most
important elements, including Fe and three other \ion{Fe}{0}-group elements, in NLTE SE.  This includes
the inclusion of thousands of lines of \ion{Fe}{1} and II in NLTE.
Something that we have {\it not} described in previous papers is that we construct our atomic
models using an automatic procedure that constructs the models from energy-level and atomic line
data in the line lists of \citet{kurucz92}.  The only input is the energy cut-off
for the highest lying levels to be included in the atomic model.  This has the very important advantage that the atomic
data for the NLTE models is bound to be consistent with that of the LTE models.  The supplementary data
for radiative bound-free ($b-f$) and collisional cross-sections that are needed are described in \citet{shorth05}. 

\paragraph{}

For the species treated in NLTE, only levels connected by
transitions of $\log gf$ value
greater than -3 (designated primary transitions) are included directly in the SE rate equations.
All other transitions of that species (designated secondary transitions) are calculated
with occupation numbers set equal to the Boltzmann distribution value with excitation
temperature equal to the local kinetic temperature, multiplied by the ground state
NLTE departure co-efficient for the next higher ionization stage.  
We have only included in our NLTE treatment here those ionization stages that are non-negligibly
populated at some depth in the Sun's atmosphere.  As a result, we only include the
first one or two ionization stages for most elements.  We therefore err on the side of including 
{\it more} ionization stages than are necessary for the late G and K class stars being modeled presently.

\paragraph{}

It is worth re-emphasizing here that our method of solving the coupled SE and radiative transfer equations is
such that the SE solution is self-consistent across all NLTE species.
For example, if transitions from two or more NLTE species overlap in wavelength, the SE solutions of the
species will be correspondingly
coupled as a natural consequence of the method.  This is significant for late-type stars in which the
spectrum is notoriously over-blanketed in the blue and near UV bands.  \citet{shorth03} and \citet{shorth05}
have studied the effect of including or excluding various groups of transitions in the NLTE SE and have 
found that the SE of the Fe-group elements has a significantly greater effect on the model structure and SED
that the that of the ``light metals''.  For this investigation, we
make no attempt to individually "hand-tune" the values of atomic parameters for particular 
transitions as one should for careful spectroscopic abundance determination.  Here, we are interested in
the {\it differential} effect on the atmospheric structure and overall SED of models as a result of
many opacity sources being treated in NLTE as compared to LTE, and our hope is that errors in the many NLTE transitions
being treated will on average approximately cancel each other out.   

\paragraph{}

We note that in NLTE mode,
PHOENIX is currently restricted to the smaller, higher quality (``small'') atomic line list discussed in
Paper I.  Therefore, the LTE models used in the comparisons here are 
those of ``Series 2'' from Paper I.  This ``small'' atomic line list consists of a 1.4 Gbyte list
adapted from lists available on Kurucz' ftp site as of 2007, {\it except} for those species
treated in NLTE, for which the line list transitions are suppressed.  For NLTE species, only
those bound-bound ($b-b$) transitions accounted for in the model atoms represented by the SE equations are  
accounted for.  The molecular line list is an 11 Gbyte file that includes all 
molecular opacity sources that are important in the Sun, among many other molecular opacity courses.  This list 
was developed for PHOENIX modeling of brown dwarfs (see, for example, \citet{helling08}) and is more
that adequate to account for molecular opacity in our coolest K stars. 

\paragraph{}

The physics of NLTE radiative equilibrium (RE) is complex in that any given $b-b$ (line) or $b-f$ (photo-ionization edge) transition 
may either heat or cool
the atmosphere when treated in NLTE with respect to LTE, depending on how rapidly the monochromatic optical
depth, $\tau_\lambda$, increases inward at the wavelength of the line or $b-f$ edge, whether the transition
falls on the Wien or the Rayleigh-Jeans side of the peak of the Planck function for the star's $T_{\rm eff}$ value,
and whether the transition is a net heater or coolant in LTE with respect to the gray atmosphere.  An understanding
of why the NLTE $T(\tau)$ structure differs from that of LTE in the way that it does would require a 
careful analysis of the role of any number of $b-b$ and $b-f$ transitions throughout the spectrum in establishing the
NLTE RE.  Such an analysis is beyond the scope of the present work.  Careful investigations of NTLE RE 
for the special case of the Sun have been carried out by \citet{anderson89} and \citet{val3}.

\paragraph{}

Fig. \ref{fatmoscomp} shows the difference in kinetic temperature of NLTE and LTE models, 
$\Delta T_{\rm Kin} = T_{\rm NLTE}(\tau)-T_{\rm LTE}(\tau)$, as a function
of continuum optical depth at 12\,000 \AA, $\tau^{\rm C}_{\rm 12000}$, 
for select models spanning the grid and showing various representative behaviors throughout the grid, 
of $T_{\rm eff}$ equal to 4000, 4750, and 5500 K, $\log g$ equal to 3.0 and 
1.0 (1.5 in the case of the 550 K model), and $[{{\rm A}\over{\rm H}}]$ values of 0.0 and -0.5.  
All models show some increase
in $T_{\rm Kin}$, by as much as $\approx$ 200 K, for $\tau^{\rm C}_{\rm 12000} \le -1$.     
For solar metallicity giants of 
$T_{\rm eff} \ge 5375$ K, this ``NLTE heating'' with respect to LTE continues to the top 
of the atmosphere.  This NLTE RE
effect has been previously found, and extensively discussed, in detailed NLTE investigations of the Sun's atmosphere
(\citet{shorth05}, \citet{anderson89}), and is caused almost entirely by the effect of NLTE on the 
Fe-group lines.  The effect is enhanced by $\approx$ 100 K near the surface in 
the atmospheres of the mildly metal poor giants.  However, for stars of $T_{\rm eff} < 5375$ K,  
the effect of NLTE is to {\it cool} the atmosphere at higher layers ($\tau^{\rm C}_{\rm 12000} \le -3$) 
by as much as $\approx$ 150 K.  Photo-ionization ($b-f$) edges in the UV of \ion{Mg}{1} ($\lambda 2514$), \ion{Al}{1} ($\lambda 2076$),
and \ion{Si}{1} ($\lambda 1682$) are transitions that are strong in most of the models throughout our $T_{\rm eff}$ range, and occur 
in a spectral region where there is still enough flux that 
they might cool the atmosphere in NLTE with respect to LTE.

\paragraph{}

Note that $\Delta T_{\rm Kin}(\tau)$ behaves erratically 
at $\tau^{\rm C} > 0$ because the $T_{\rm Kin}(\tau)$ structure steepens in the lower atmosphere 
where many radiative transitions become optically thick and the evaluation of $\Delta T_{\rm Kin}$ 
becomes numerically sensitive to this slope.  
However, this is also the $\tau^{\rm C}$ range in which convection rather than radiation 
increasingly determines the $T_{\rm Kin}(\tau)$ structure as $\tau^{\rm C}$ increases, and is not as 
useful for assessing the effect of NLTE on the RE $T_{\rm Kin}$ structure.

\subsection{Synthetic spectra \label{smod_synspec}}

For both LTE and NLTE models we computed self-consistent synthetic spectra in the
$\lambda$ range 3000 to 8000 \AA~ with a spectral resolution 
($R={\lambda\over\Delta\lambda}\approx 350\,000$)
to ensure that spectral lines were adequately sampled.  We note that the value of $\xi_{\rm T}$ was
consistent between the spectrum synthesis and the input atmospheric models, as was all the
stellar parameters.  In the NLTE calculations, PHOENIX also
automatically adds additional $\lambda$ points to adequately sample the spectral lines
that correspond to $b-b$ atomic transitions that are being treated in NLTE.  
These were then degraded to match the low resolution measured $f_\lambda$ distributions of B85
by convolution with a Gaussian kernel of FWHM value equal to 75 \AA.  This is about three times the 
nominal sampling, $\Delta\lambda$, of 25 \AA~ claimed by B85, and we found that it provided
the closest match to the appearance of the B85 spectra, as discussed in Paper I.  
We note that this convolution also automatically accounts approximately for macro-turbulence, which 
has been found to be around 5.0 km s$^{\rm -1}$ for G and K II stars \citep{gray82}.  
We interpolate in $\log f_\lambda$ between adjacent synthetic SEDs to obtain a SED grid
with an effective sampling, $\Delta T_{\rm eff}$, of 62.5 K.  The accuracy of this interpolation
was investigated in Paper I, and was found to be accurate to within 5\% in {\it linear} flux among the coolest
models where the variation in $f_\lambda$ with $T_{\rm eff}$ is greatest.  This is about the same, 
or smaller, than the typical $\Delta T_{\rm eff}$ value between
adjacent spectral subclasses for GK stars. 

\subsubsection{NLTE}

\begin{figure}
\plotone{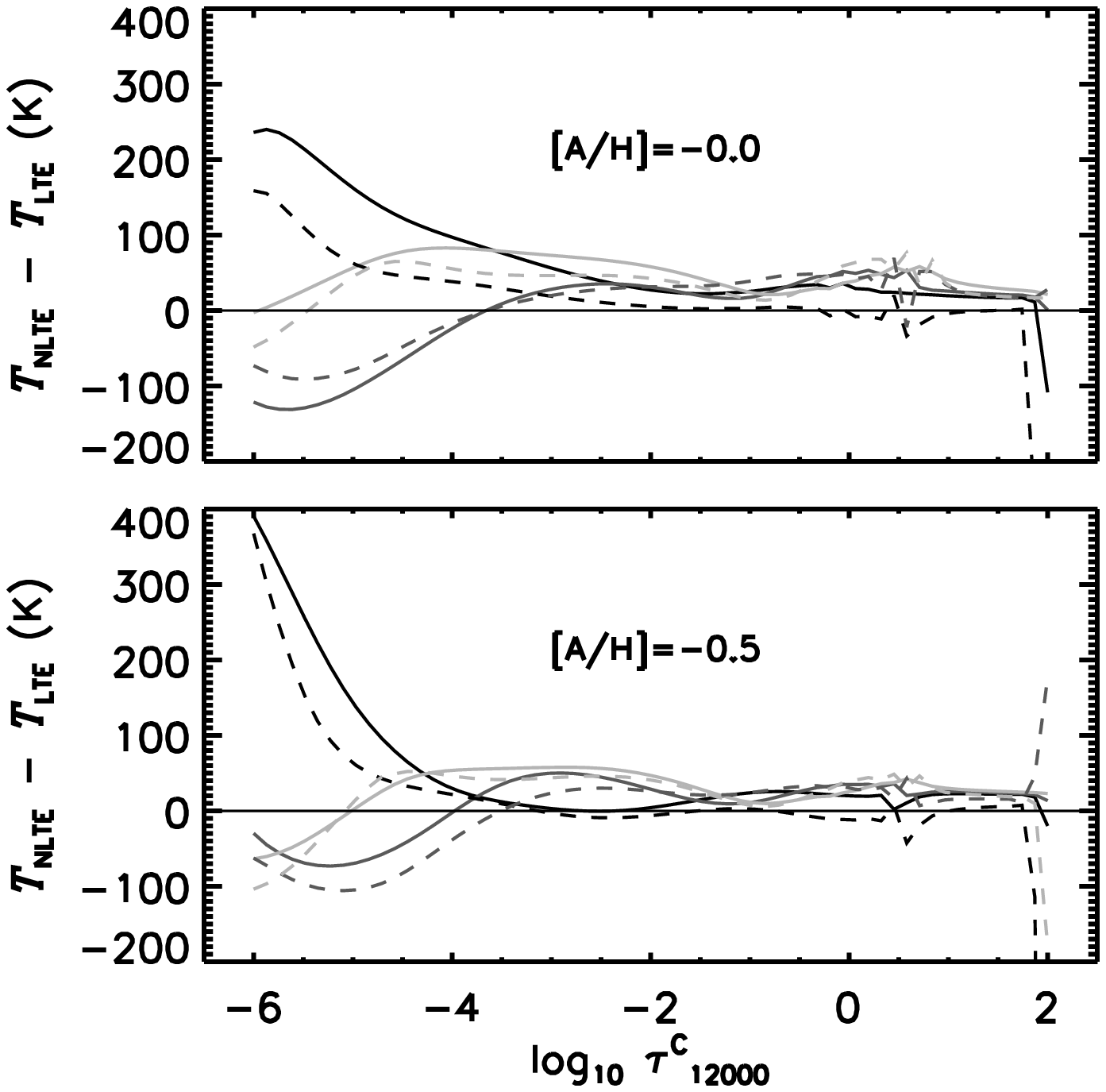}
\caption{
The difference in kinetic temperature of select NLTE and LTE models, 
$\Delta T_{\rm Kin} = T_{\rm NLTE}(\tau)-T_{\rm LTE}(\tau)$.  
The $x$-axis: Continuum optical depth at 12\,000 \AA, $\tau^{\rm C}_{\rm 12000}$. 
Results are shown for models of $T_{\rm eff}$ of 5500 K (black line), 4750 K 
(medium gray line) and 4000 K (light gray line) at $\log g$ values of 3.0 (solid line)
and 1.0 (or 1.5 for the 5500 K model) (dotted line).  
Upper panel: $[{{\rm A}\over{\rm H}}]=0.0$; Lower panel: $[{{\rm A}\over{\rm H}}]=-0.5$.  
  \label{fatmoscomp}}
\end{figure}

\begin{figure}
\plotone{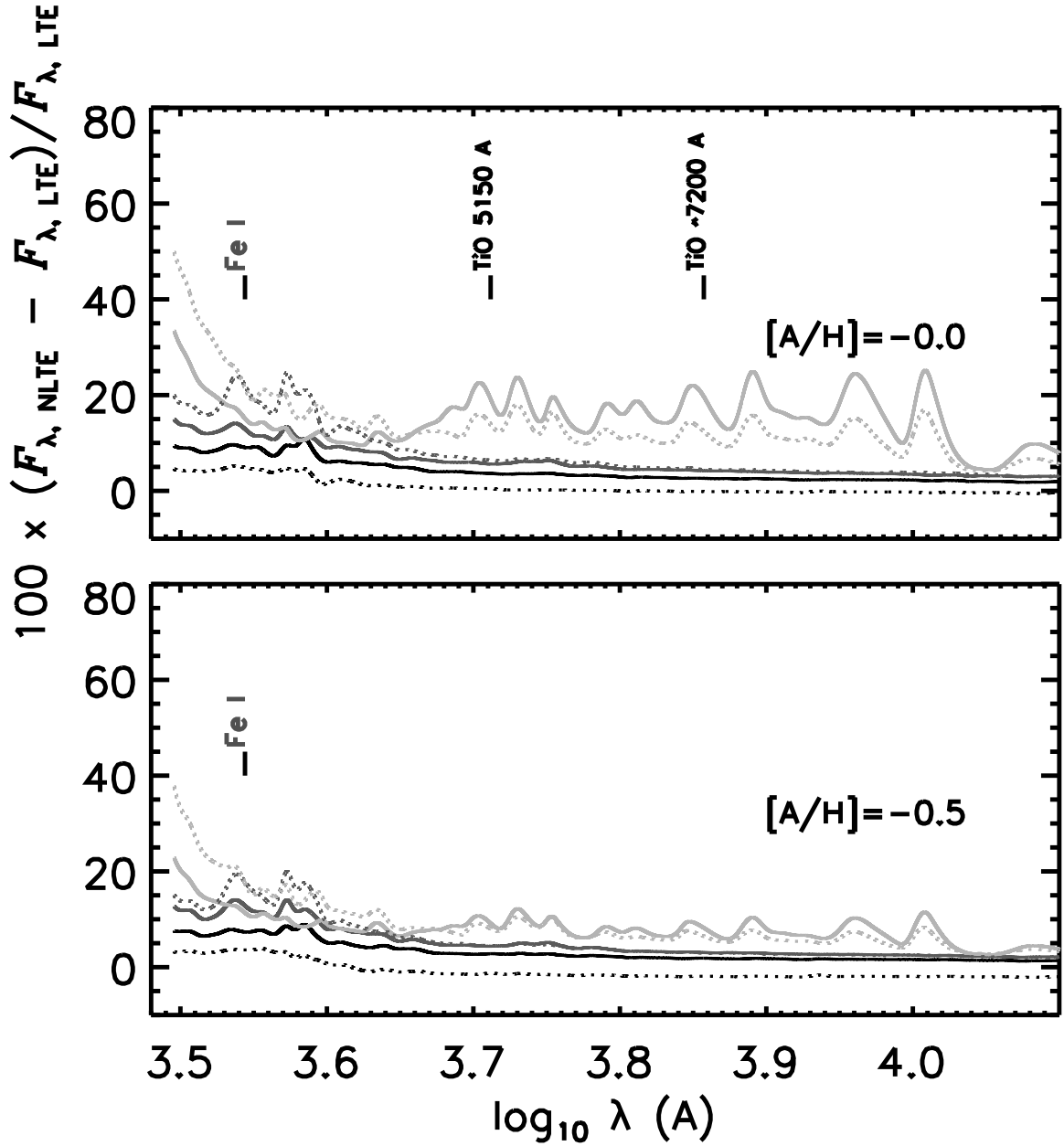}
\caption{
The same as Fig. \ref{fatmoscomp}, but showing the relative difference in $f_\lambda$ of the convolved 
NLTE and LTE synthetic SEDs as a percentage, 
$100.\times(f_\lambda, {\rm NLTE}-f_\lambda{\rm LTE})/f_\lambda{\rm LTE}$.  For clarity, we have convolved
the relative difference to the ~75 \AA~ effective resolution of the observed spectra.
  \label{fspeccomp}}
\end{figure}

\begin{figure}
\plotone{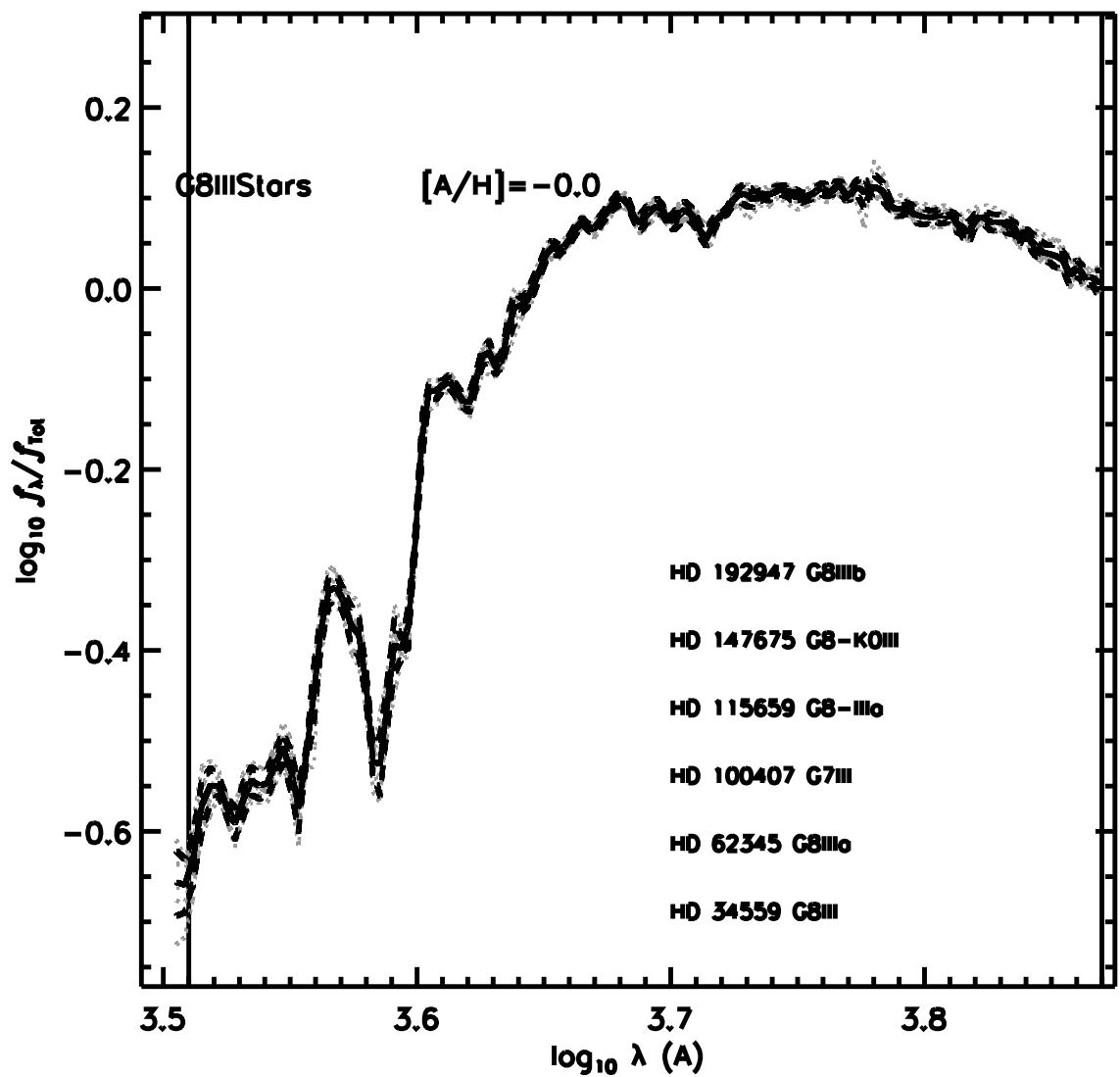}
\caption{G8 III/$[{{\rm A}\over{\rm H}}]=0.0$ sample.  Eight individual normalized spectra, 
$\log f_{\lambda}/f_{\rm Tot}$ (see text), of six stars from the B85 catalog
 that met our quality-control criteria (gray dotted lines).
Sample average spectrum: black solid line; $\pm 1~ \sigma$ spectra: black dashed lines. 
Vertical lines near the ends of the $x$-axis range show the $\lambda$ limits
of the ``quasi-bolometric'' normalization area (see text).\label{fvarymult}}
\end{figure}

Fig. \ref{fspeccomp} shows the relative difference of the NLTE and LTE synthetic SEDs, 
$100. \times (f_{\lambda, {\rm NLTE}}-f_{\lambda{\rm LTE}})/f_{\lambda{\rm LTE}}$, convolved to the effective resolution 
of the observed SEDs (75 \AA) for the models of Fig. 
\ref{fatmoscomp}.   Generally, the NLTE SEDs become increasingly
brighter than the LTE SEDs as $\lambda$ decreases.  This is a well-known effect that has been studied
extensively in the Sun (see \citet{rutten86}, \citet{anderson89}) and is caused by the NLTE over-ionization
(really, LTE {\it under}-ionization!) of the minority \ion{Fe}{1} stage.  The NLTE effect on the \ion{Fe}{1}/II 
ionization equilibrium reduces the extinction in the ``forest'' of \ion{Fe}{1} lines that blanket the spectrum
(the ``iron curtain'')
and allows more flux to escape.  Because the lines are more densely concentrated per unit $\Delta\lambda$
as $\lambda$ decreases, the blue and near-UV bands are effected significantly more than the red band. 
This effect dominates any change in $f_\lambda$ that might be expected from the NLTE effect on
the $T_{\rm Kin}$ structure that is seen in Fig. \ref{fatmoscomp}. 
As a result, we expect that fitting NLTE SEDs to observed SEDs would lead to a {\it lower} inferred
$T_{\rm eff}$ value.  For the coolest models in the grid ($T_{\rm eff} \le 4125$ K), 
the NLTE spectra are also brighter in the regions of
strong molecular bands, such as that of TiO around $\log\lambda=3.86$, as a result of the outer atmosphere
being warmer in NLTE (see Fig. \ref{fatmoscomp}) and less favorable to molecule formation.  
As a result, we expect that fitting either the ratio of the blue- to red-band flux, or the
strength of the molecular bands, would lead to a lower $T_{\rm eff}$ value when using NLTE models
as compared to LTE models.

\paragraph{}

The synthetic SEDs were 
interpolated to the same regular $\lambda$ grid as that of the processed 
B85 spectra, and the same 
``quasi-bolometric'' normalization was applied (see section \ref{sobsseds}).
This normalization differs from the single-point
normalization used in Paper I, and has the advantage of not biasing the
fit of model to observed spectra to any particular wavelength.  

\paragraph{}

\begin{figure}
\plotone{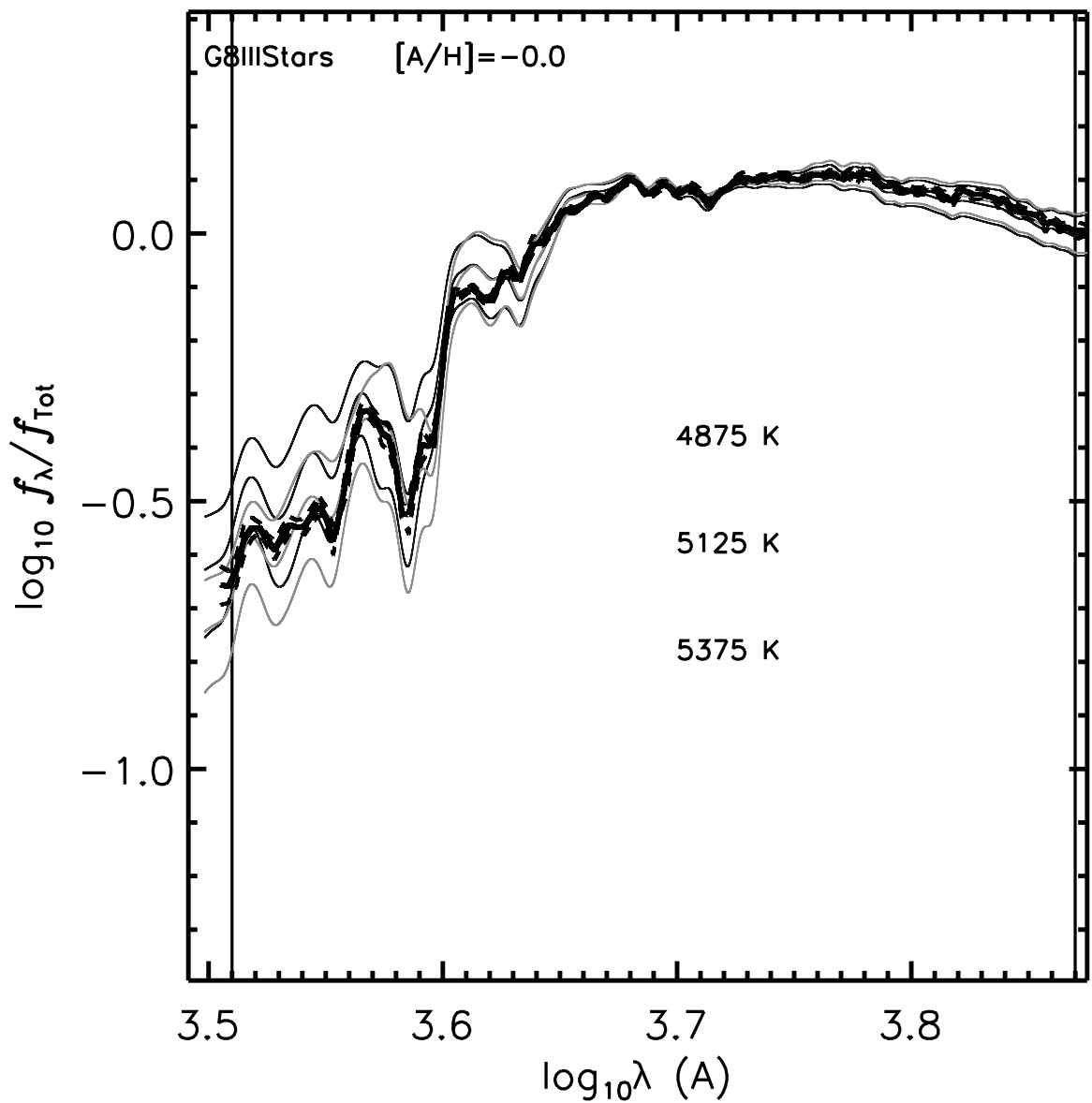}
\caption{G8 III sample: Comparison of normalized observed sample average to select normalized synthetic $f_{\lambda}$ spectra
of NLTE models.  
Thick solid line: sample average $f_{\lambda}$ spectrum, black dashed lines: 
$\pm 1~ \sigma$ spectra.  Thin solid gray-scale lines: select synthetic $f_{\lambda}$ spectra among those bracketing the model of
best fit $T_{\rm eff}$ value at the smallest and largest $\log g$ values of the model grid; Dark gray: 
$\log g=3.0$, light gray: $\log g=1.0$.    
  \label{fcompG8III}}
\end{figure}

\begin{figure}
\plotone{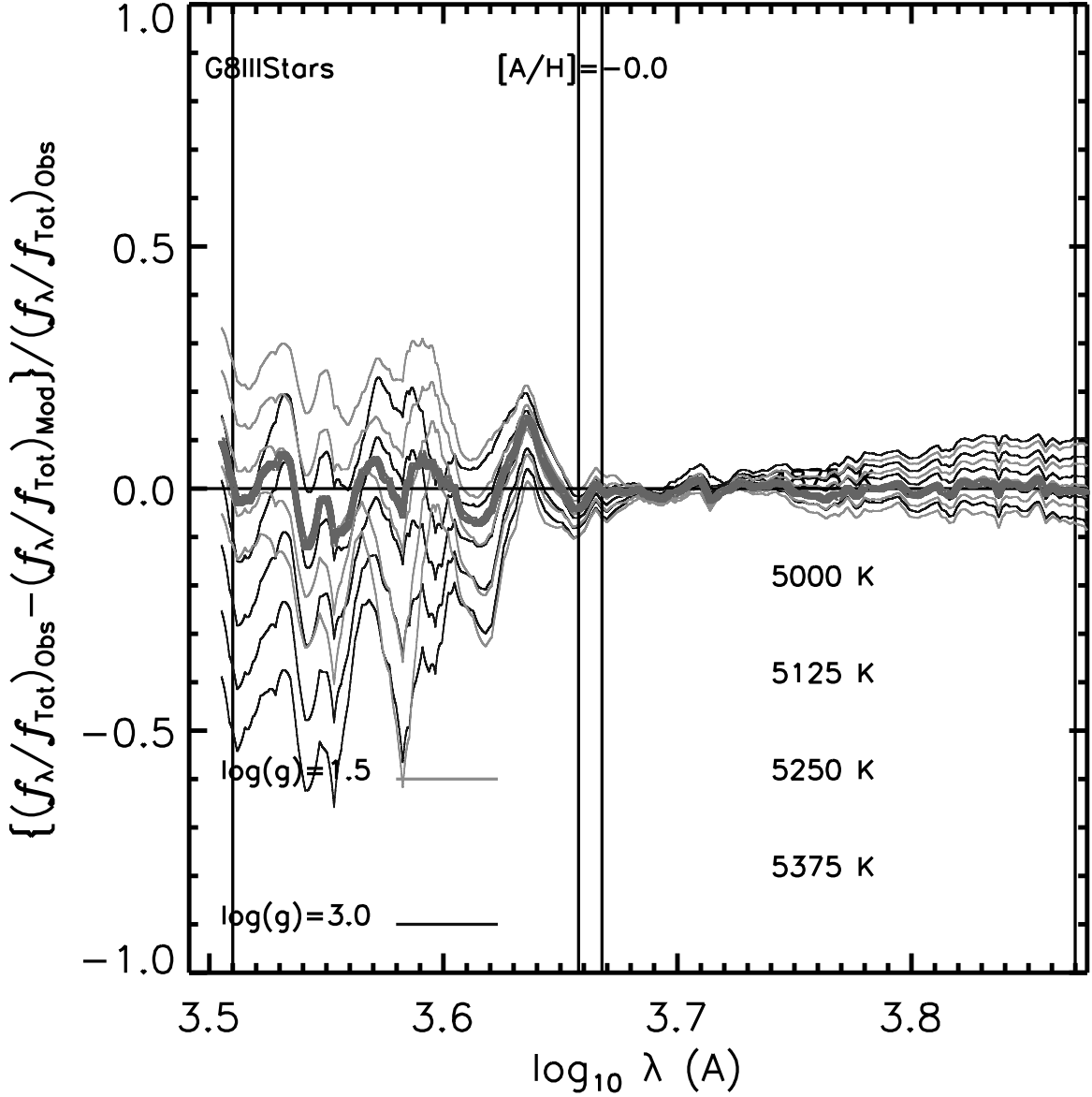}
\caption{G8 III sample - Relative difference between the observed normalized sample average 
$f_{\lambda}$ spectrum and select normalized NLTE synthetic $f_{\lambda}$ spectra among those bracketing the 
model of best fit $T_{\rm eff}$ value, at the smallest and largest $\log g$ values of the model grid.  
The horizontal line
indicates a difference of zero.  The vertical lines represent the boundaries of the ``blue'' and
``red'' bands.  Thick line: closest matching synthetic $f_{\lambda}$ spectrum. 
  \label{fdiffG8III}}
\end{figure}

As an illustrative example, Fig. \ref{fcompG8III} shows the comparison of the mean and
$\pm 1 \sigma$ spectra of the observed $f_{\lambda}$ distributions with 
a selection of
NLTE synthetic $f_{\lambda}$ distributions for models bracketing the best fit $T{\rm eff}$ value at
the smallest and largest $\log g$ values in the model grid for the G8 III/$[{{\rm A}\over{\rm H}}]=0.0$ sample.
Fig. \ref{fdiffG8III} shows the {\it difference}
between the mean of the observed $f_{\lambda}$ distribution and a selection of 
NLTE synthetic distributions for models bracketing the best fit $T{\rm eff}$ value at
the smallest and largest $\log g$ values in the model grid, relative to the observed mean distribution,
$( f_{\lambda, {\rm Mean~ Observed}} - f_{\lambda, {\rm Model}} ) / f_{\lambda, {\rm Mean~ Observed}}$
for the same sample.  We note that Paper I shows similar comparisons for the LTE
synthetic spectra for a variety of samples.

\section{Goodness of fit statistics \label{fitstats}}

We compute on the interpolated $\lambda$ grid for each spectral class sample the root mean 
square {\it relative} 
deviation, $\sigma$, of the mean observed $f_{\lambda}$ distribution from
the closest matching and bracketing convolved synthetic $f_{\lambda}$ 
distributions in the $\lambda$ 
range from 3200 to 7000 \AA, according to\\

\begin{equation}
\sigma^2 = {1\over N}\sum_i^N ((f_{\lambda, {\rm Obs}}-f_{\lambda, {\rm Mod}})/f_{\lambda, {\rm Obs}})^2
\end{equation}

 where $N$ is the number of $\lambda$ points in the $\lambda$ grid in the 3200 to 7000 \AA~ range.  
We also compute separate RMS values, $\sigma_{\rm Blue}$ and $\sigma_{\rm Red}$, for our 
nominal ``blue'' and ``red'' sub-ranges
of 3200 to 4600 \AA~ and 4600 to 7000 \AA, respectively.  A comparison of
the $\sigma_{\rm Blue}$ and $\sigma_{\rm Red}$ values indicates how well the synthetic
spectra fit in the blue and near UV band given the quality of fit in the red band.
A break-point of 4600 \AA~ was chosen on the basis of visual inspection of
where the deviation of the synthetic from the observed spectrum starts to become rapidly larger as $\lambda$ decreases.

\paragraph{}
 
In Tables \ref{tabstats1} and \ref{tabstats2} we present the $\sigma$, $\sigma_{\rm Blue}$, 
and $\sigma_{\rm Red}$ values for the LTE and NLTE models, respectively, along with the
best fit value of $T_{\rm eff}$ and $\log g$ for each star.  The value of the model 
$[{{\rm A}\over{\rm H}}]$ is also tabulated, although, its value was specified {\it a priori}
on the basis of the metallicity catalog of \citet{cayrelsr01} rather than fitted.  As a check, we also
computed $\sigma$ values for each sample with the $[{{\rm A}\over{\rm H}}]$ values of the 
models reversed; {\it ie.} we fitted models of $[{{\rm A}\over{\rm H}}]=0.0$ to samples
formed from stars of catalog $[{{\rm A}\over{\rm H}}]$ equal to $\approx -0.5$ and
{\it vice versa}.  In most cases the $\sigma$ values of the metal-reversed fits were 
larger, and in a few cases were comparable to, those of the original fits.  In no cases
were they lower.  We conclude that the $[{{\rm A}\over{\rm H}}]$ values of \citet{cayrelsr01}
are generally reliable for GK stars to within $\pm \approx 0.25$.  (For those stars within our 
spectral class range for which the catalog gives an uncertainty estimate, usually taken from the sources 
they are citing, their estimates range from 0.03 to 0.09.  For Arcturus (K1.5 III), with 17 measurements, 
and HD62509 (K0 III) with seven measurements, the RMS deviations of $[{{\rm A}\over{\rm H}}]$ are 0.111
and 0.086, respectively.)

\subsection{Trend with $T_{\rm eff}$}

\begin{figure}
\plotone{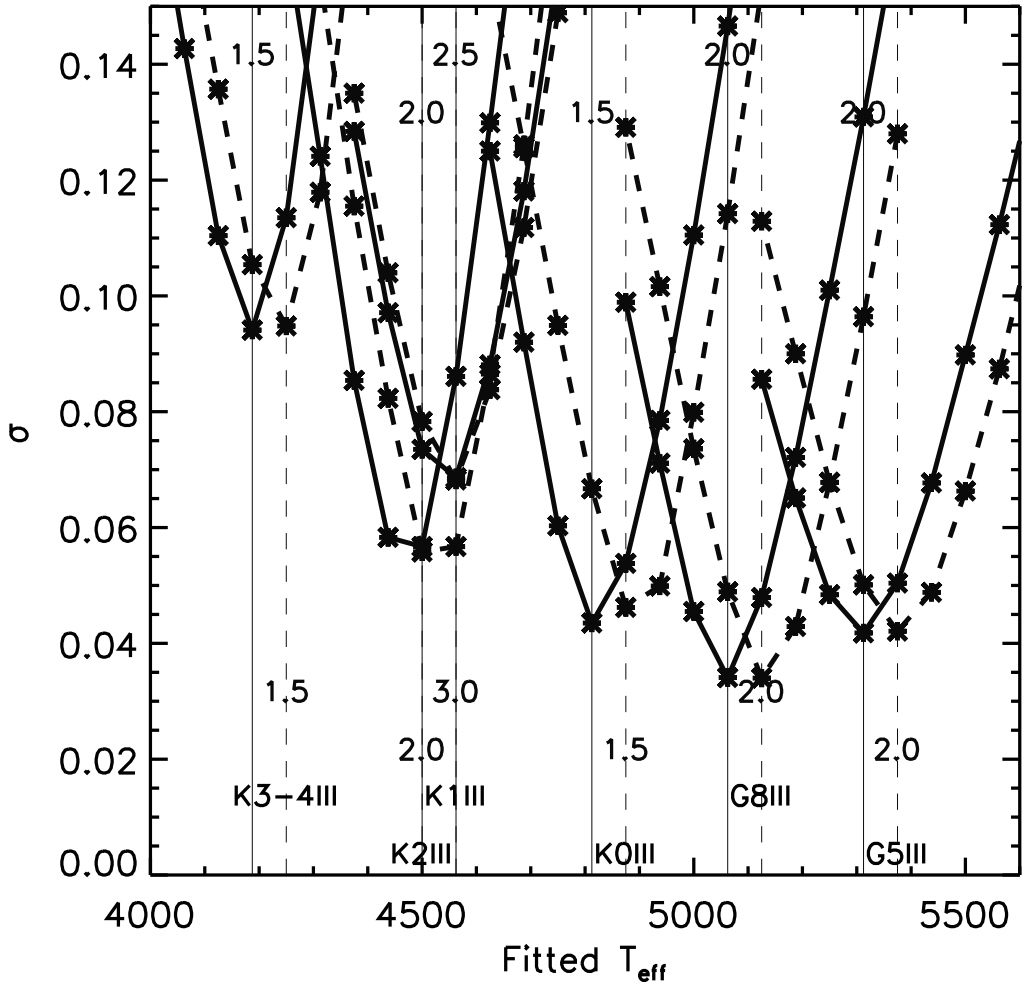}
\caption{Giants of solar metallicity: Variation of $\sigma$ with model $T_{\rm eff}$.   
Solid line: NLTE models; dashed line: LTE models.  Vertical lines: Best fit $T_{\rm eff}$
values.  Best fit $\log g$ values are given for the LTE (lower row), 
and NLTE (upper row) models. 
\label{fstatsgnt}}
\end{figure}

Fig. \ref{fstatsgnt} shows the variation of
$\sigma$ with model $T_{\rm eff}$ ($\sigma(T_{\rm eff})$ curves)
for the giant stars of solar metallicity, for both LTE and NLTE models.  The $\log g$ 
value of the best fit model for each spectral class is also indicated.  
The quality of the fit generally worsens with decreasing $T_{\rm eff}$, as is seen by the increase of $\sigma_{\rm Min}$ 
for later spectral class.
As noted in Paper I, the density of spectral lines generally increases with increasing lateness.
Therefore, this trend in the discrepancy between synthetic and observed SEDs could be
explained by inadequacies in the input atomic data for bound-bound ($b-b$) transitions,
or by inadequacies in the treatment of spectral line formation.  Moreover, spectral features, especially those 
of molecules, are very sensitive to 3D effects 
\citet{asplund00}, and that also contributes to increasing discrepancies for the cooler models.  

\paragraph{}

Interestingly, we note that the $\sigma_{\rm Min}$ values
for the LTE and NLTE models differ negligibly from each other for all spectral classes.  
The adoption of NLTE does not improve the quality of fit provided by the best fit model. However, the value of 
the best fit NLTE $T_{\rm eff}$ is always one $\Delta T_{\rm eff}$ element (62.5 K) lower than the LTE value for 
giants of any spectral class.
This was expected from the comparison of the LTE and NLTE $f_\lambda$
distributions in Section \ref{smod_synspec}, and amounts to a uniform shift downward in the
$T_{\rm eff}$ calibration of the GK III classification by $\approx$ 62.5 K.  Unfortunately,
because the shift is one $\Delta T_{\rm eff}$ element, we are barely resolving the shift numerically,
and the actual shift could be anywhere in the range of about 30 to 95 K, and could vary with 
spectral class within this range.  
For all six spectral classes (Tables \ref{tabstats1}, \ref{tabstats1}), we find best fit $\log g$ values from NLTE modeling in the range of 1.5 to 2.5.  For the 
LTE models the variation in best fit $\log g$ values is larger, with the K 1 III sample yielding a value of 3.0, which
is near the upper limit for early K III stars.  This may be taken a marginal evidence that the NLTE models provide more
physically realistic parameters.

\subsubsection{Red {\it vs} blue band}   

\begin{figure}
\plotone{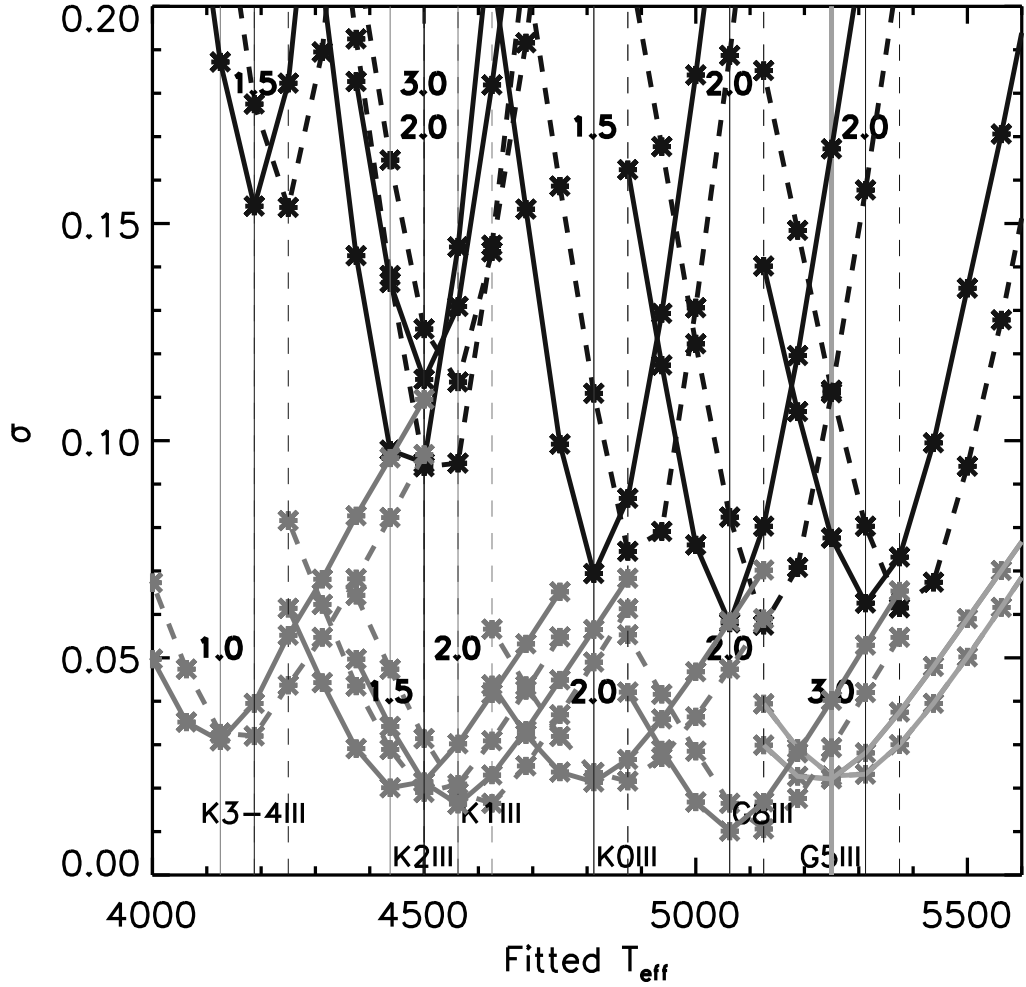}
\caption{Same as Fig. \ref{fstatsgnt}, but for the separate fits to the blue (dark gray lines) and red 
(light gray lines) bands.  The red band results for the G5 III sample are highlighted in medium gray to indicate
that they are problematic (see text).  \label{fstatsgntrb}}
\end{figure}

The quality of the best fit, as indicated by the value of $\sigma_{\rm Min}$, 
rapidly deteriorates for spectral classes later than K0 (Fig. \ref{fstatsgnt}).  This is not unexpected; as 
$T_{\rm eff}$ decreases,
the SED becomes increasingly line blanketed, particularly in the blue band, and the quality of the fit
is increasingly dependent on the quality of atomic data and the treatment of line formation. 
Correspondingly, from Fig. \ref{fdiffG8III} it can be seen that the difference
spectra show increasing variability around the zero line as $\lambda$ decreases, 
in addition to any systematic trend away from the zero line.  
This can be seen more directly in Fig. \ref{fstatsgntrb}, which shows the variation with $T_{\rm eff})$ of $\sigma$ 
for the blue ($\sigma_{\rm Blue}(T_{\rm eff})$) and red ($\sigma_{\rm Red}(T_{\rm eff})$ bands
separately. 
For samples of spectral class K0 and warmer, the $\sigma{\rm Red}, Min$ value is lower than the $\sigma_{\rm Blue}, Min$ 
value by $\approx 0.05$ because the longer $\lambda$ range is less complicated by
line blanketing.  This discrepancy between $\sigma_{\rm Blue, Min}$, and $\sigma_{\rm Red, Min}$ increases rapidly for later spectral classes. 
We note that for {\it all} spectral classes, the -62.5 K $\Delta T_{\rm eff}$ offset between best fit NLTE and LTE models 
is also found 
separately in the blue and red bands.

\paragraph{}

For the G8 and K0 III $[{{\rm A}\over{\rm H}}]=0.0$ spectral classes (the special case G5 III is discussed separately in 
Section \ref{sG5III}), 
the red and blue bands yield the same best fit value of $T_{\rm eff}$.  This consistency 
across wave bands that have very different amounts of line blanketing provides some assurance of the 
the quality of the modeling, but does not distinguish the quality of the NLTE treatment from that of the LTE.  
For the K1 III $[{{\rm A}\over{\rm H}}]=0.0$ sample the best fit $T_{\rm eff}$ value found from the blue band is
one $\Delta T_{\rm eff}$ element (62.5 K) cooler than that found from the red band.  This may indicate that
the NLTE treatment over-estimates the amount of NLTE blue-band 
$f_\lambda$ brightening (discussed is Section \ref{smod_synspec}), thus leading to an artificially low $T_{\rm eff}$ value with 
respect to the less blanketed red band.  This is consistent with the results of \citet{shorth09} for Arcturus (K1.5 III). 
However, for the K2 and K3-4 III $[{{\rm A}\over{\rm H}}]=0.0$ samples, the best fit $T_{\rm eff}$ value found from the blue band is
62.5 K {\it hotter} than that found from the red band, indicating that
for the most heavily line blanketed giants considered here, the NLTE treatment {\it under}-estimates the amount of NLTE blue-band 
$f_\lambda$ brightening (discussed is Section \ref{smod_synspec}), thus leading to an artificially high $T_{\rm eff}$ value with 
respect to the less blanketed red band.  

\paragraph{}

For the metal-poor giant samples the situation is also confused: For the G8 III $[{{\rm A}\over{\rm H}}]=-0.5$ sample
the $T_{\rm eff}$ value derived from the blue band is 62.5 K hotter than that from the red band, whereas it is 62.5 K
cooler in the case of the K1.5 III $[{{\rm A}\over{\rm H}}]=-0.5$ sample (Arcturus).  We note that for the case of
Arcturus, for which the observed spectra presumably have the best quality, the use of NLTE models reduces the size of
the $T_{\rm eff}$ discrepancy between blue and red bands from 125 to 62.5 K.  

\paragraph{}

The lack of any clear trend between
the sign of the blue- and red-band $T_{\rm eff}$ results and spectral class most likely is a reflection of the lack of
good fit in the blue band provided by {\it any} model.  Any signal in the value of the fitting statistic indicating how well 
any model fits at those wavelength windows where the fit is good is diluted by the ``noise'' from all the wavelength 
windows where all models, including the best fit one, are grossly discrepant with the observations.

\subsubsection{G5 III sample \label{sG5III}}

As noted in Paper I, the behavior of the variation of the $\sigma_{\rm Red}(T_{\rm eff})$ curve for the G5 III stars
is peculiar and leads to a spurious result for the best fit value of $T_{\rm eff}$.  From
Fig. 7 of Paper I it can be seen that this is caused by a broad absorption feature exhibited
by the observed SED with respect to the model SEDs ranging from a $\log\lambda$ value of 3.753 to 3.774
(5660 to 5940 \AA).  As a result, the value of $\sigma_{\rm Red}$ is increased significantly,
even for models that provide a good match to the overall spectrum.  Therefore, our
best fit value of $T_{\rm eff}$ for the G5 III models is best determined from 
the blue band alone.  This deficit of absorption in the synthetic SEDs with respect to the
observed ones is consistently present in the individual observed spectra for
the G5 III stars, spans 12 data points in the raw observed spectrum, and varies smoothly with
wavelength over a range of ~280 \AA.  
  We note that this
discrepancy is either absent, or much less pronounced, in both the G4-5 V and G8 III stars, so
appears to be localized in both $T_{\rm eff}$ and $\log g$.  In Paper I we compared our three G5 III
spectra from the B85 catalog with spectra for G4 and G6 III stars in the stellar spectrophotometric library of 
\citet{jacobyhc84} and concluded that this discrepancy is likely caused by a data acquisition or calibration
error in the B85 data. 


\subsection{Trend with $[{{\rm A}\over{\rm H}}]$ and $\log g$}

\begin{figure}
\plotone{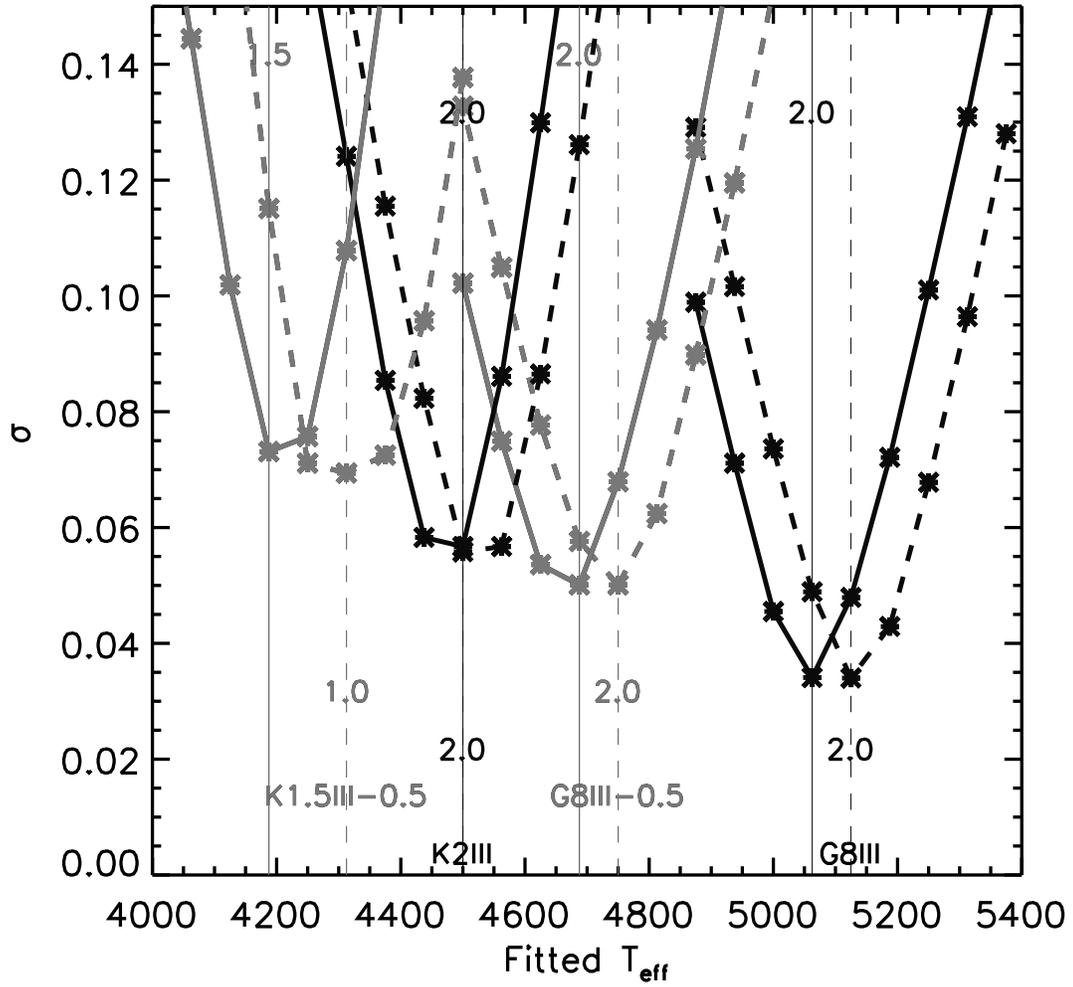}
\caption{Same as Fig. \ref{fstatsgnt}, but for the metal poor giants (lighter lines), along with select solar metallicity giants
of the same (or similar) spectral class for comparison (darker lines).  
  \label{fstatsmtl}}
\end{figure}

\begin{figure}
\plotone{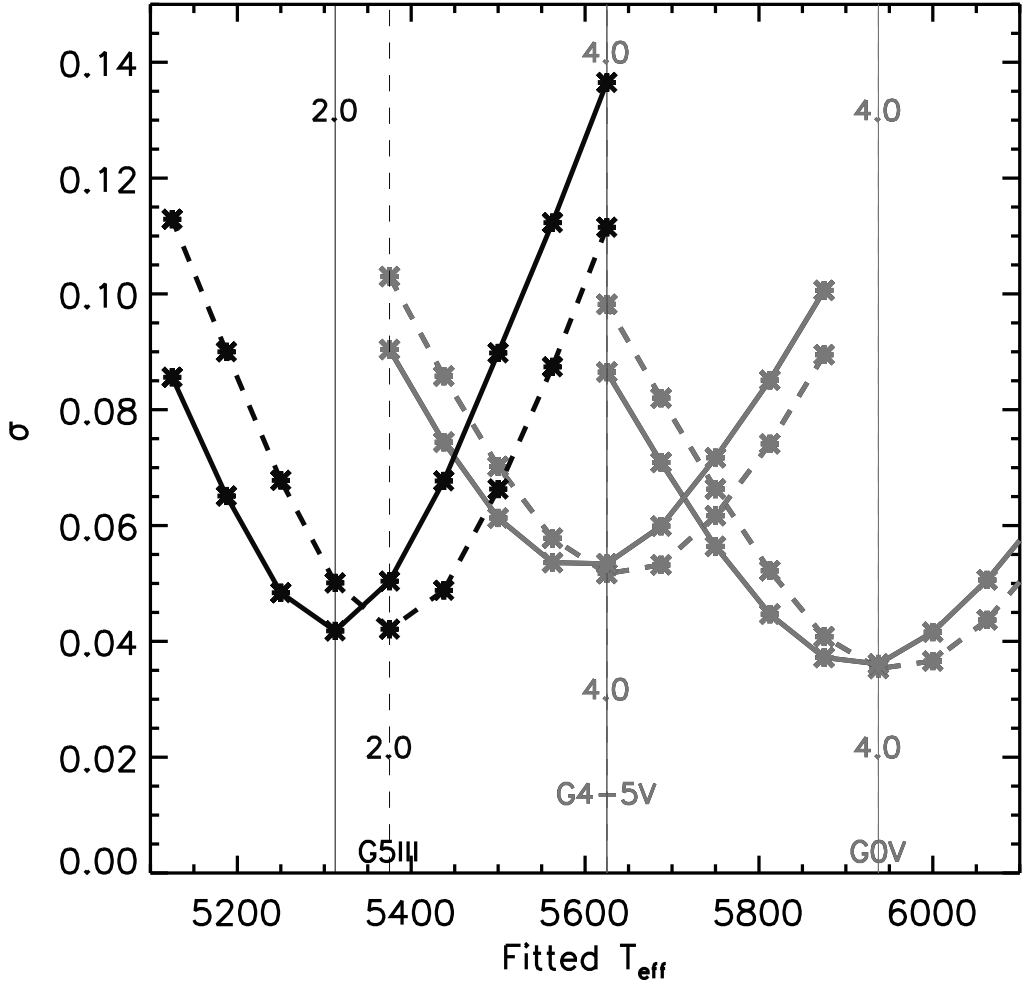}
\caption{Same as Fig. \ref{fstatsgnt}, but for the dwarf stars (lighter lines), along with the G5 III sample (darker lines) for comparison
to the G5 V sample.  
The blue band fit for the G5 III sample ($\sigma_{\rm Blue}(T_{\rm eff}$)) is shown, rather than the whole visible band fit (see text).
  \label{fstatsdwf}}
\end{figure}

Figs. \ref{fstatsmtl} and \ref{fstatsdwf} show the $\sigma(T_{\rm eff})$ curves for the whole band fits
for the G0 and G4-5 V ($[{{\rm A}\over{\rm H}}]=0.0$) samples, and the giant samples of 
$[{{\rm A}\over{\rm H}}]=-0.5$ (G8 and K1.5), respectively.  
Also shown are $\sigma(T_{\rm eff}$) curves
for select giants of $[{{\rm A}\over{\rm H}}]=0.0$ for comparison.   
Because of 
the special problem of the red band in our G5 III sample (discussed in Section \ref{sG5III}),
we show the result of the G5 III fit in the blue band in Fig. \ref{fstatsdwf}.  The best fit parameters for the whole 
band, and for the red and blue bands are also given in Tables \ref{tabstats1} and \ref{tabstats2}. 

\paragraph{}

For the metal poor giants (Fig. \ref{fstatsmtl}), the results are qualitatively similar
to those for the solar metallicity giants: NLTE models give minimum $\sigma$ values
that are effectively the same as those of LTE models at each spectral class, and the 
NLTE grid yields best fit $T_{\rm eff}$ values that are one $\Delta T_{\rm eff}$ element
cooler that those of LTE grid. 
We note that at a given spectral class, the quality of fit ($\sigma_{\rm Min}$ value) is worse
as $[{{\rm A}\over{\rm H}}]$ decreases, which may initially seem unexpected if 
the treatment of line blanketing is the greatest obstacle to achieving a good match.
However, we note that $T_{\rm eff}$ is correlated with $[{{\rm A}\over{\rm H}}]$ 
at fixed spectral class ({\it eg.} both the G8III/-0.5 and K1.5III/-0.5 samples are
300 to 350 K cooler than the G8III/0.0 and K2III/0.0 samples, respectively) and so the 
real trend is likely to be the same correlation between $\sigma_{\rm Min}$ and
$T_{\rm eff}$ that was seen for solar metallicity giants Fig. \ref{fstatsgnt}.    

\paragraph{}

For the dwarfs (Fig. \ref{fstatsdwf}), the LTE and NLTE models yield the same best fit values of $T_{\rm eff}$.
However, the $\sigma(T_{\rm eff})$ curves are flatter, and those of the NLTE models are skewed
toward lower $T_{\rm eff}$ than those of the LTE models.  For both dwarf spectral classes, 
$\sigma(T_{\rm eff}(\sigma_{\rm Min})-1 \Delta T_{\rm eff})$ is approximately the same as 
$\sigma(T_{\rm eff}(\sigma_{\rm Min}))$. 
We infer that for class V stars, the NLTE reduction
in the value of $T_{\rm eff}$ also exists, but that is it $\approx 0.5 \Delta T_{\rm eff}$
({ie.} $\approx$ 31 K).

\subsection{Arcturus}

  From a comparison of Tables \ref{tabstats1} and \ref{tabstats2} for the Arcturus sample (K1.5 III, $[{{\rm A}\over{\rm H}}]=-0.5$) 
the $T_{\rm eff}$ value from the LTE blue band fit is 125 K lower than that from the red band, whereas with the NLTE modeling,
it is only 62.5 K lower.  That the NLTE grid yields $T_{\rm eff}$ values that are more consistent across wave bands provides some evidence
that these models are more realistic.  However, that there is still a discrepancy at all indicates that our NLTE
models may be over-estimating the blue $f_\lambda$ level, and hence leading to an artificially low $T_{\rm eff}$ value,
 with respect to the red band.
This is
consistent with the results of \citet{shorth03} and \citet{shorth09}, who also compared
models to the observed $f_\lambda$ distribution of B85.  Note that this is the opposite to what was found for
the solar metallicity K2 III sample, so the effect may be metallicity dependent.

\section{Comparison to other $T_{\rm eff}$ calibrations}



%
%
%
%

In Paper I we compared our various LTE $T_{\rm eff}$ values to the less model dependent 
calibrations from the infrared flux method (IRFM) of \citet{ramirezm05} (RM05), and from
interferometric angular diameters of K giants determined with the CHARA array \citet{baines10} (B10), 
with a brief summary 
of these calibrations, a justification for these comparisons, and a discussion of
how we interpolated or extracted appropriate $T_{\rm eff}$ values for comparison.  We note
that RM05 and B10 estimate their $T_{\rm eff}$ values to be accurate to $\pm 75$ K and 50 to 150 K (2 - 4\%),
respectively.   
Here we choose to compare our NLTE
results to RM05 and B10 again, along with recently derived
$T_{\rm eff}$ values for large samples of G and early K giants from three additional sources.
 \citet{wang11} derived $T_{\rm eff}$ values for 99 G-type giants by requiring the $[{{\rm Fe}\over {\rm H}}]$ values  
derived from \ion{Fe}{1} lines in spectra acquired with the  
High Dispersion Spectrograph (HDS, $R=60\, 000$) at the Subaru Telescope in the 4900 - 7600 \AA~ range to be
independent of the excitation energy of the lower level ($\chi_{\rm l}$).  $[{{\rm Fe}\over {\rm H}}]$ values
are derived from the equivalent widths, $W_\lambda$, of Fe lines.  They also independently
re-derived $T_{\rm eff}$ from photometric $T_{\rm eff}$ relations of \cite{alonsoam01} and reddening laws in the literature combined
with catalog values of a number of photometic indices.  For the latter they estimate an uncertainty of 
$\pm\approx 100$ K from the $T_{\rm eff} - B-V$ relation.  They note that the $T_{\rm eff}$ values from the 
\ion{Fe}{1} lines are on average $44 \pm 117$ K larger than those derived from photometric calibrations. 
\citet{takeda08} used ATLAS9 atmospheric models to derive atmospheric parameters and $[{{\rm Fe}\over {\rm H}}]$ values from
 the $W_\lambda$ values of \ion{Fe}{1} and II lines
in the 5000 - 6200 \AA~ region of 322 bright ($V < 6$) late-G giants with spectra ($R=67\, 000$) obtained with the
HIDES spectrograph at the 1.88 m telescope of the Okayama Astrophysical Observatory.  They determine statistical
uncertainties in their $T_{\rm eff}$ values of 10 - 30 K.
\citet{mishenina06} used line depth ratios (from 70 to 100 ratios per star) to determine $T_{\rm eff}$ values for ~200 
late-G and early-K clump giants with spectra in the 4400 - 6800 \AA region ($R \approx 42\,000$) from the ELODIE 
echelle spectrograph at the 1.93 m telescope of the Haute-Provence Observatoire.  They determine that the $1 \sigma$
uncertainties are 5 to 25 K.  They also determine $[{{\rm Fe}\over {\rm H}}]$ values from the $W_\lambda$ values of
\ion{Fe}{1} lines while requiring that all \ion{Fe}{1} and II lines yield the same abundance to fix $\log g$ and $\xi_{\rm T}$.  
For the latter three studies, we extracted stars for which the derived $[{{\rm Fe}\over {\rm H}}]$ value was within
0.1 of either of the two $[{{\rm A}\over {\rm H}}]$ values of our model grid (0.0 and -0.5).

\paragraph{}  

\begin{figure}
\plotone{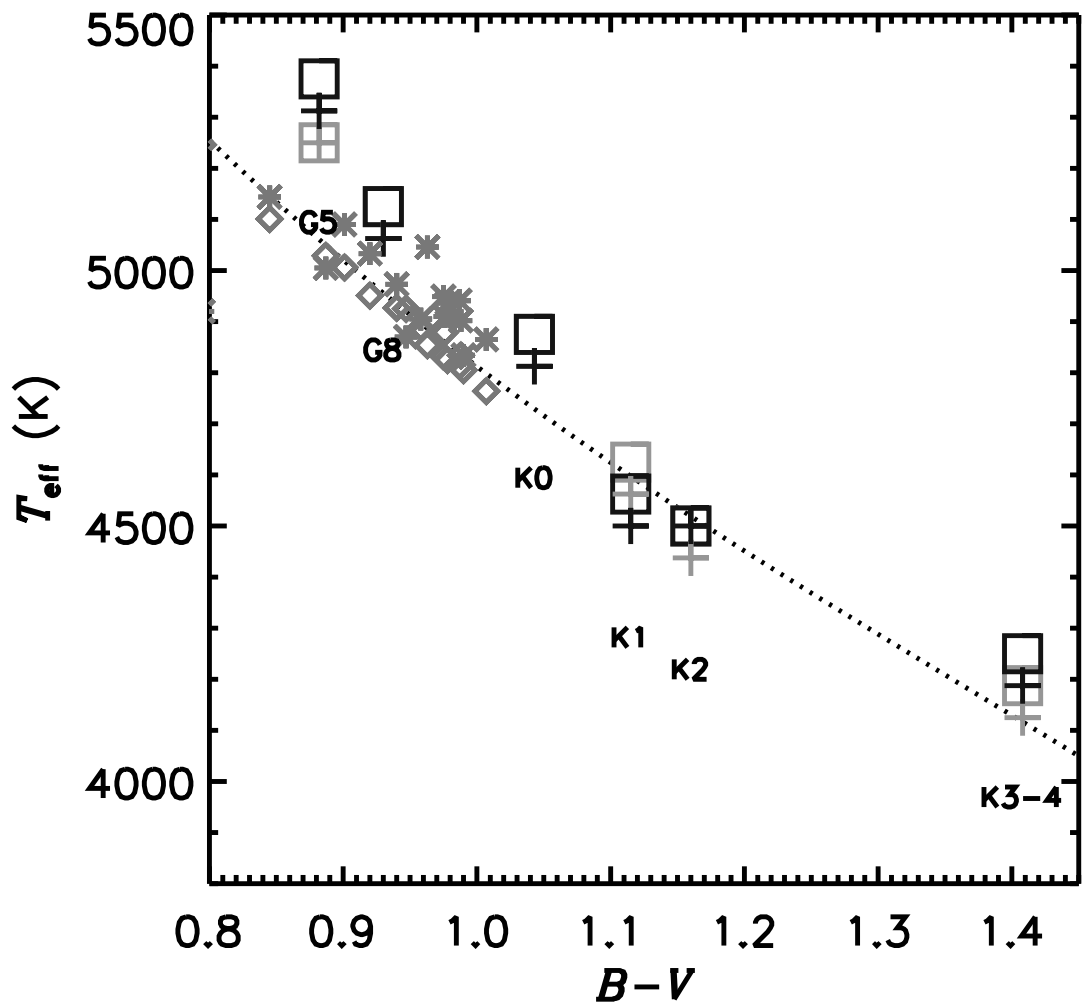}
\caption{Solar metallicity giants: Comparison of our best fit $T_{\rm eff}$ values with various calibrations of $T_{\rm eff}$ 
as a function of $B-V$.
Squares: LTE models; Crosses: NLTE models.  Black symbols: Fit to the blue band; Gray symbols: 
fit to the red band.  Note that for some cases the red and blue band symbols exactly overlap.  Calibrations of RM05 
(dotted line), \citet{wang11} photometric
(diamonds), and \citet{wang11} spectroscopic (asterisks). 
  \label{fcalibgnt1}}
\end{figure}

\begin{figure}
\plotone{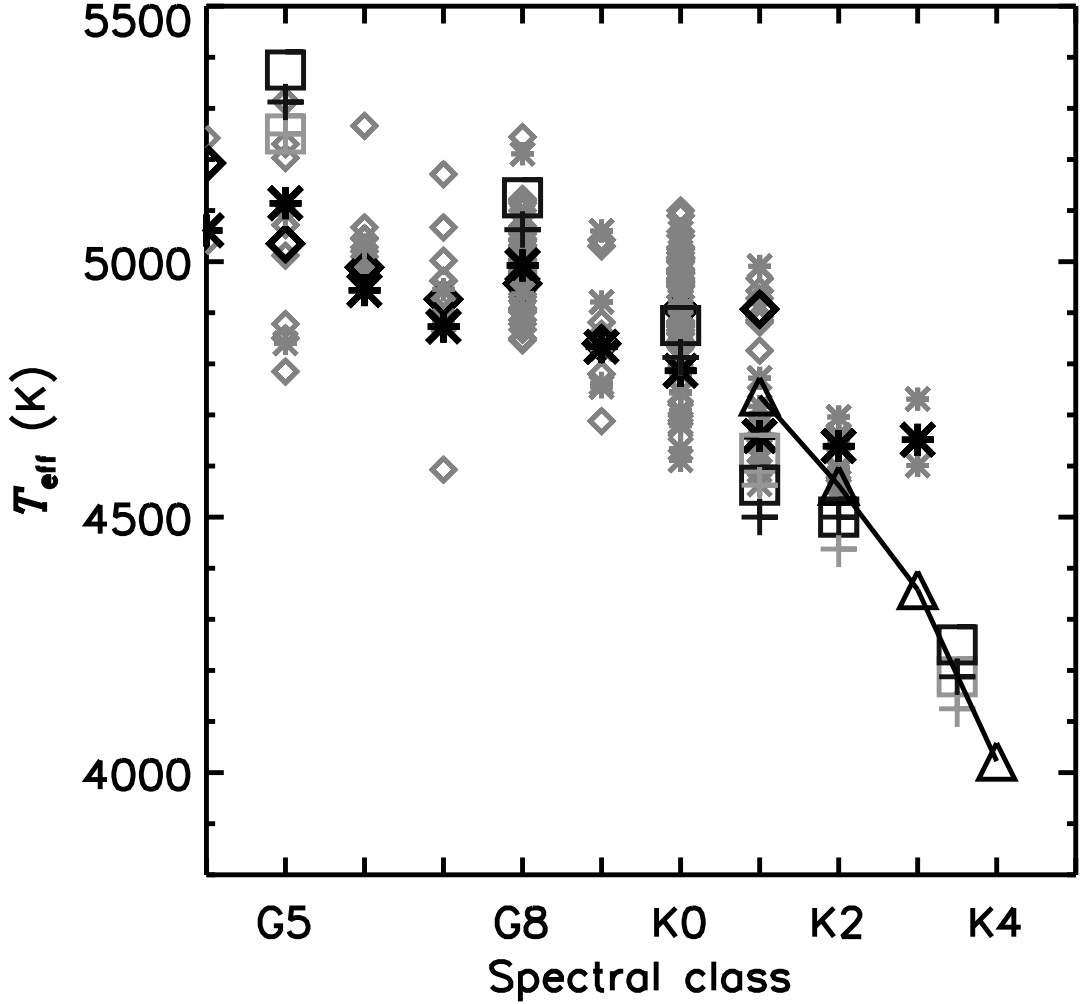}
\caption{Same as Fig. \ref{fcalibgnt1}, but for various calibrations of $T_{\rm eff}$ 
as a function of spectral class.
Calibrations of B10 (solid line with triangles), \citet{takeda08} (diamonds), and 
\citet{mishenina06} (asterisks).  For the \citet{takeda08} and \citet{mishenina06} results, the larger
black symbols are averages, weighted by number of stars, computed by us. 
  \label{fcalibgnt2}}
\end{figure}

\begin{figure}
\plotone{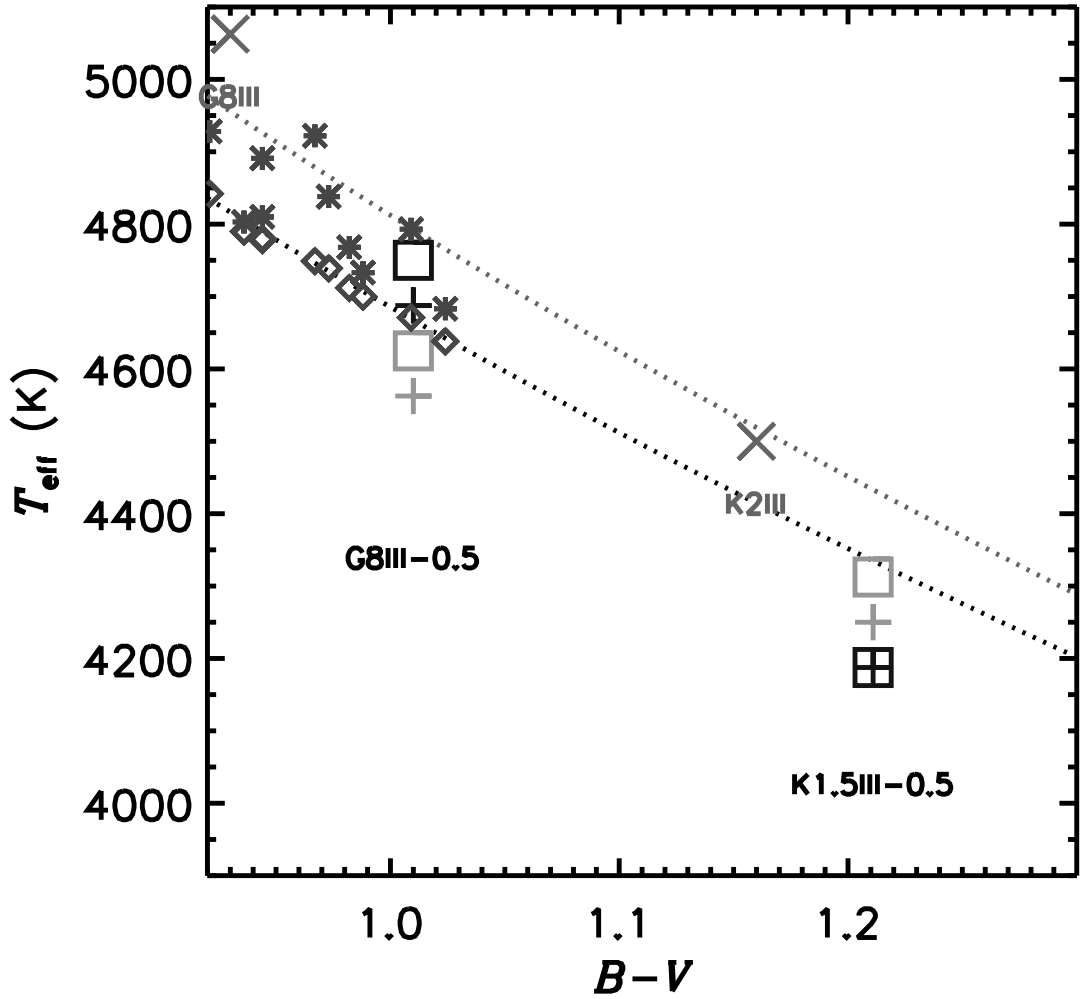}
\caption{Same as Fig. \ref{fcalibgnt1}, but for the metal poor giants.  (Note: B10 does not provide
a $T_{\rm eff}$ calibration for metal poor giants.)  For comparison we also show the RM05 calibration 
for solar metallicity giants (lighter dotted line), and our NLTE results for the solar metallicity G8 and
K2 III samples.   
\label{fcalibmtl}}
\end{figure}

One point worth reiterating from Paper I is that the RM05 calibration is especially useful
because it spans a wide range of values of $B-V$ and $[{{\rm A}\over{\rm H}}]$ at luminosity
classes V and III.  Therefore, we are able to compare all our results to RM05.  We have extracted from the published tables of \citet{wang11}, \citet{takeda08}, and \citet{mishenina06}
samples of giants with $-0.1 < [{{\rm A}\over{\rm H}}] < 0.1$ and $-0.6 < [{{\rm A}\over{\rm H}}] < -0.4$
for comparison to our results for our $[{{\rm A}\over{\rm H}}]=0.0$ and $-0.5$ samples,
respectively.  
In Paper I, to facilitate the comparison, we computed mean and RMS ($\sigma$)
$B-V$ values for each of our spectral class samples using colors for individual objects 
from the Catalog of Homogeneous Means in the UBV System \citep{mermilliod91}.  We use the 
same mean colors for our samples here.  
In Table \ref{tabcomp1} and Figs. \ref{fcalibgnt1} through \ref{fcalibmtl} we 
present a comparison of our 
$T_{\rm eff}$ values fitted to our blue and red spectral ranges, and those of the
RM05 and B10 calibrations.  We note from Fig. \ref{fcalibgnt1} that the photometrically derived
$T_{\rm eff}$ values of \citet{wang11} agree very closely with the calibration of RM05.  This is
expected because \citet{wang11} and RM05 both make use of the photometric index {\it versus}
$T_{\rm eff}$ relations of \cite{alonsoam01} (and papers in that series).

\subsection{Solar metallicity giants}

Our LTE models match the RM05 calibration to within the precision of the grid ($\rm 62.5$ K) for the
latest spectral classes, and increasingly predict too large a $T_{\rm eff}$ value, by as much as
$\approx 300$ K as $B-V$
decreases.  This seems surprising given that the later-type stars have more complicated SEDs that
are more difficult to model, as discussed above.  This may reflect of a ``conspiracy'' of canceling
errors at the latest spectral classes, and the result should be approached with caution.  
  Again, we caution that the red band result for the G5 III sample is spurious for the reasons discussed 
above.  Recently, \citet{casagrande10} have published a new IRFM $T_{\rm eff}$ scale (with
I. Ramirez and J. Melendez as co-authors) for stars of $\log g > 3.0$ and find that the scale is
warmer by ~85 K than that of RM05 for stars of $T_{\rm eff} \ge \approx 5000$ K while agreeing more
closely with RM05 for cooler stars.  This deviation from the RM05 scale results from a change in the 
absolute calibration of the photometry, therefore, it is expected to also apply to lower gravity stars. 
If it does, then our results may be in closer agreement with the IRFM calibration across the whole range
of spectral classes studied here.     
Our LTE results are in similarly close agreement to the K giant $T_{\rm eff}$ calibration of B10.   
Because our NLTE $T_{\rm eff}$ scale is 62.5 K lower than the LTE scale, the NLTE models
predict too low a $T_{\rm eff}$ value for the latest types, and a value that is closer to that of RM05,
but still to large, for the earlier types.  
We note that the results of B10 are based on limb-darkening derived from 1D atmospheric models,
and that \cite{chiavassa10} recently found that limb-darkening from 3D models leads to smaller 
derived radii and $T_{\rm eff}$ values that are correspondingly larger by as much as ~20 K for 
stars of $T_{\rm eff}$ in the range 4600 to 5100 K (spectral classes K0 to G5) and 
$[{{\rm A}\over {\rm H}}]$ of -1, and by a smaller
amount for stars of $[{{\rm A}\over {\rm H}}]$ of 0.0.  It is intriguing that the sign of the 3D 
correction is one that would bring the B10 results into closer agreement with our NLTE $T_{\rm eff}$ 
values for the corresponding spectral classes. 

\paragraph{}

We note that the $T_{\rm eff}$ values for individual stars  
derived by \citet{wang11}
from the \ion{Fe}{1}/II balance are also generally larger than RM05, and are in closer agreement with 
our values.  Because our method is also essentially spectroscopic rather than photometric, this
might seem to be evidence for spectroscopic $T_{\rm eff}$ determinations being generally 50-100 K larger
than photometric determinations.  However, the photometric $T_{\rm eff}$ scale is dependent upon the 
absolute calibration adopted, and we caution against drawing a conclusion on the basis of this work.  
The $T_{\rm eff}$ values for individual stars of \citet{takeda08} and
\citet{mishenina06} for G giants show a significant scatter and our $T_{\rm eff}$ values lie near the
upper limit of their results.  We have computed star-count weighted means of their $T_{\rm eff}$
values and also show them in Fig. \ref{fcalibgnt2}.  Our values for G giants are larger than this mean trend,
as was found for our comparison to the RM05 calibration.

\subsection{Solar metallicity dwarfs and metal poor giants}

  RM05 is the only calibration we have to compare our results to for class V stars.  Our results for G dwarfs are
better than those for for G giants, 
in that for both the G0 and G4-5 samples 
our blue band NLTE $T_{\rm eff}$ values are just slightly warmer than the RM05 calibration, by one $\Delta T_{\rm eff}$
element (62.5 K).  For G stars, we infer that our NLTE modeling is increasingly accurate as $\log g$ increases.  
This may reflect that our 1D horizontally homogeneous static models become increasingly inaccurate as $\log g$
decreases. 

\paragraph{}

Our LTE $T_{\rm eff}$ value for the K1.5 III sample of $[{{\rm A}\over{\rm H}}]=-0.5$ (consisting entirely of Arcturus 
spectra, recall)
provides about the same quality of match to the RM05 calibration as that of our K1 and K2 III samples of 
$[{{\rm A}\over{\rm H}}]=0.0$.  At the same time,
our LTE $T_{\rm eff}$ value for the metal poor G8 III sample is much closer to the RM05 calibration than that of the solar 
metallicity G8 III sample.  The NLTE blue band fit at G8 III/$[{{\rm A}\over{\rm H}}]=-0.5$ is very close RM05, whereas
the NLTE results are cooler than RM05 by $\approx 150$ K at K1.5 III/$[{{\rm A}\over{\rm H}}]=-0.5$.  We tentatively infer 
that our ability to reproduce the RM05 calibration with NLTE models for the earlier GK spectral classes improves with 
decreasing metallicity in this $[{{\rm A}\over{\rm H}}]$ range.  This is not unexpected given the decreasing dependence
on the realism of the line blanketing treatment as $[{{\rm A}\over{\rm H}}]$ decreases.

\section{Conclusions}

Our strongest conclusion is that the adoption of NLTE for many opacity sources 
shifts the spectrophotometrically determined $T_{\rm eff}$ scale for giants downward by
an amount, $\Delta T_{\rm eff}$, in the range of about 30 to 90 K all across the
mid-G to mid-K spectral class range, and across the $[{{\rm A}\over{\rm H}}]$ range from
0.0 to -0.5.  
This shift brings our 
spectrophotometrically derived $T_{\rm eff}$ scale for the solar metallicity G giants into closer agreement
with the less model-dependent $T_{\rm eff}$ scale determined by the IRFM, although our $T_{\rm eff}$ 
values for these G giants are too large in any case.  For the K giants, LTE and NLTE models
provide about the same quality of match, and are closer to the less model-dependent IRFM $T_{\rm eff}$ 
values than is the case for the G giants. 
We find tentative evidence on the basis of two spectral classes in the G range that this NLTE 
downward shift in the $T_{\rm eff}$ scale becomes smaller as luminosity class increases 
from III to V. 

\paragraph{}

  Both NLTE and LTE model SEDs show a much greater variation about the observed SED in our 
more heavily line blanketed ``blue'' band ($\lambda < 4600$ \AA) than in the red band.  
This probably indicates that there are inadequacies in the accuracy and completeness of the atomic 
line list data and in the treatment of line formation.  The latter inadequacy may in part be 
a result of our use of static 1D models.  Nevertheless, we find somewhat surprising agreement in 
the best fit value of $T_{\rm eff}$ between the blue and red bands.  There is marginal evidence that NLTE models
seem to give more consistent results between the blue and red bands for the earlier
spectral classes (G8-K0) of solar metallicity than for later classes.  Moreover, there is marginal evidence that
the derived $\log g$ values are more consistent between the red and blue bands from NLTE modelling than that of LTE.

\paragraph{}

Presumably, the highest quality observed SED in library is that for the 
K1.5III/$[{{\rm A}\over{\rm H}}]\approx -0.5$ sample, which consists of three independent 
measurements of the spectrum of the bright standard red giant Arcturus.  We find that our 
NLTE grid provides greater consistency in derived $T_{\rm eff}$ value between our blue
and red bands than does the LTE grid.  However, we find that the blue band yields a 
$T_{\rm eff}$ value that is still lower than that of the red band (by nominally 62.5 K),
indicating that NLTE models of red giants predict too much flux in the blue band with respect 
to the red band.  This is a recurrence of a long-standing problem with the modeling of
late-type stellar SEDs (see \citet{shorth09}), and may indicate an inadequacy in the 
atmospheric modeling of such stars.  However, we do not find strong evidence of
this blue band versus red band discrepancy among our many solar metallicity SED fits,
and speculate that it may be a discrepancy that worsens with decreasing metallicity.

\paragraph{}

As a by-product of this investigation, we have produced a quality-controlled stellar library of
observed mean and $\pm 1 \sigma$ SEDs for solar metallicity giants that well sample the range from G8 to K4 III.
We will make both the library of observed SEDS and the NLTE (and corresponding LTE) grid of model
SEDS available to the community by ftp \\
(http://www.ap.smu.ca/ ~ishort/PHOENIX).

\subsection{Future directions}

 That no model provides a good fit for many wavelength windows in the blue band suggests that 
a more sophisticated statistical test of goodness-of-fit, in which the contribution at each wavelength
to the statistic is weighted by the ability of any model to provide a fit at that wavelength.  We plan to 
investigate statistical tests that might enhance the signal of agreement, of lack thereof, between 
the red and blue bands for any model.

\paragraph{}   

 The $griz$ photometric system employed in large surveys such as that 
of the Sloan Digital Sky Survey (SDSS) have become increasingly important for the characterization
of late-type stars (see, for example, the exhaustive analysis of \citet{pinsonneault11} that was 
made public just as we were drafting this report).  It would be useful to investigate whether 
synthetic colors computed from our model SEDs in this,
and possibly other intermediate band photometric systems optimized for stellar photometry, 
are sensitive to NLTE effects.  We plan to expand our NLTE grid by incorporating
non-solar abundance distributions for metal poor populations (mainly $\alpha$-enhancement) and
much lower metallicities typical of the halo population.  Very metal poor halo giants are
important tracers of the Galaxy's early chemical evolution, and the effect of a {\it large scale} 
NLTE treatment, such as that performed here, on their derived parameters and compositions has yet to 
be carried out.

\acknowledgments

CIS is grateful for NSERC Discovery Program grant 264515-07.  The calculations were
performed with the facilities of the Atlantic Computational Excellence Network (ACEnet).

\clearpage




\clearpage


\clearpage

\clearpage

\clearpage

\clearpage


\clearpage

\clearpage

%

\clearpage

\clearpage

\clearpage

\clearpage

\clearpage

\clearpage

\clearpage

\clearpage

\begin{figure}
\plotone{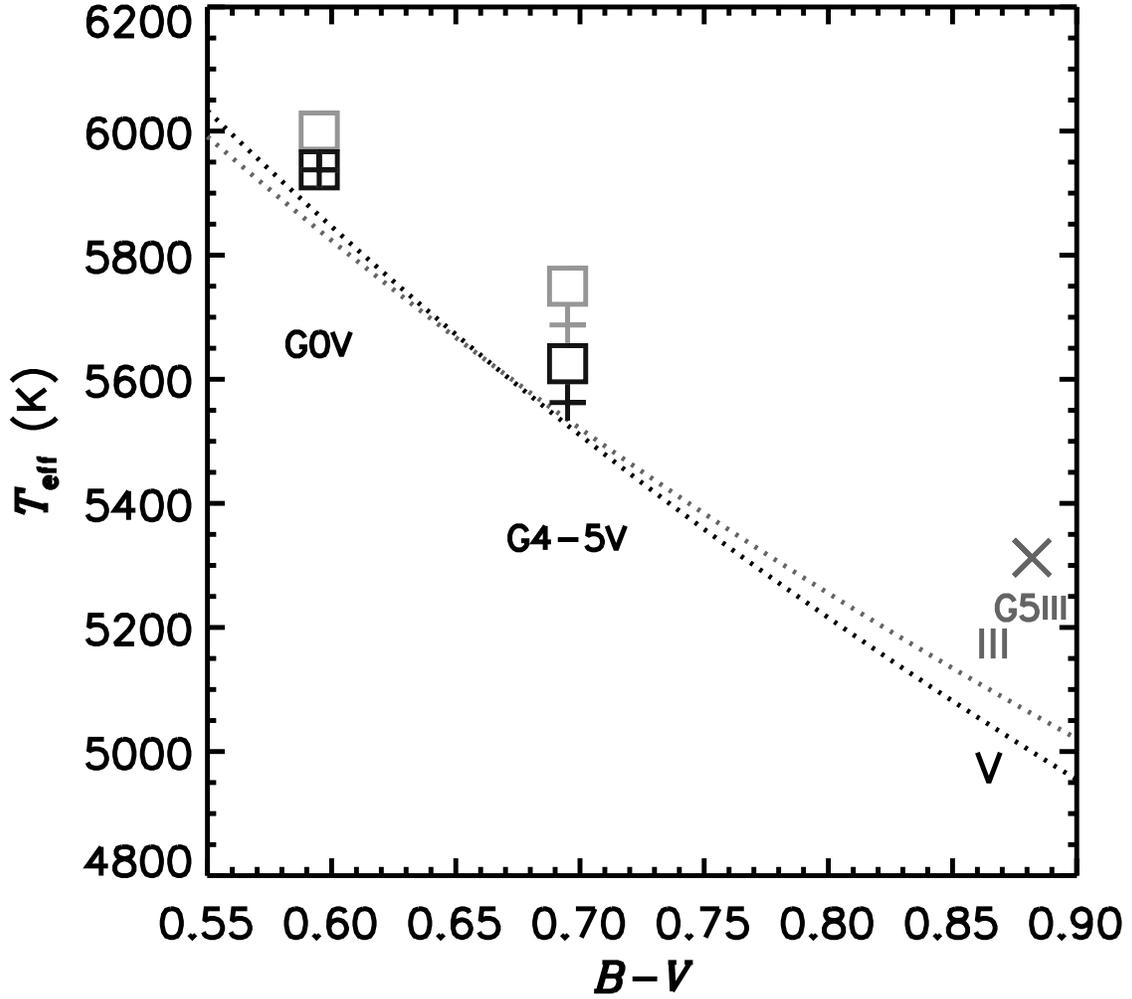}
\caption{Same as Fig. \ref{fcalibgnt1}, but for the dwarfs, calibration of RM05 only.  
For comparison we also show the RM05 calibration for giants (lighter dotted line), and our blue band
NLTE result for the G5 III sample. \label{fcalibdwf}}
\end{figure}






\clearpage

\begin{deluxetable}{lllll}
\tablecolumns{5}
\tablecaption{List of spectral class/$[{{\rm A}\over{\rm H}}]$ samples, with the number of stars used to 
form each sample, the mean (and RMS) $B-V$ value of the stars comprising the sample (from \citep{mermilliod91}), 
the total number of $[{{\rm A}\over{\rm H}}]$ values in the \citet{cayrelsr01} catalog among the stars 
comprising the sample, and the number of individual spectra used to form each sample (the entry in column 5 
is larger than that of column 2 when one or more stars in the sample has more than one independenet spectrum
 in the B85 catalog).} 
\tablehead{
\colhead{Spectral} & \colhead{Num}   & \colhead{Mean}           & \colhead{Num}                       & \colhead{Num} \\ 
\colhead{type}     & \colhead{stars} & \colhead{$B-V (\sigma)$} & \colhead{[${{\rm A}\over{\rm H}}$]} & \colhead{spectra} 
} 
\startdata
G5 III & 2 & 0.882 (0.019) & 3 & 3 \\
G8 III & 6 & 0.930 (0.004) & 8 & 8 \\
K0 III & 10 & 1.043 (0.002) & 27 & 14 \\
K1 III & 3 & 1.115 (0.003) & 3 & 3 \\ 
K2 III & 2 & 1.160 (0.006) & 3 & 3 \\
K3-4 III & 4 & 1.408 (0.014) & 5 & 4 \\
\hline
G0 V & 2 & 0.595 (0.004) & 11 & 2 \\
G5 V & 2 & 0.695 (0.010) & 7 & 2 \\
\hline
G8 III & 2 & 1.010 (0.000) & 4 & 3 \\
K1.5 III\tablenotemark{a} & 1 & 1.211 (0.009) & 17 & 3\\
\enddata
\tablenotetext{a}{Arcturus, $\alpha$ Boo}
\label{tabb85}
\end{deluxetable}

\clearpage

%
\begin{deluxetable}{lrrrrrrrrrr}
\tablecolumns{11}
\tablecaption{LTE models: Closest match models to mean sample spectra and goodness of fit statistics. }
\tablehead{
\colhead{}              & \multicolumn{3}{c}{Total SED}      &  \multicolumn{3}{c}{Blue}                     & \multicolumn{3}{c}{Red}                      & \colhead{} \\ 
\colhead{Spectral type} & \colhead{$T_{\rm eff}$} & \colhead{$\log g$} & \colhead{$\sigma_{\rm Min}$} & \colhead{$T_{\rm eff}$} & \colhead{$\log g$} & \colhead{$\sigma_{\rm Min}$} & \colhead{$T_{\rm eff}$} & \colhead{$\log g$} & \colhead{$\sigma_{\rm Min}$} & \colhead{$[{{\rm A}\over{\rm H}}]$} \\
} 
\startdata
G5 III & 5375$^{\rm a}$ & 2.0 & 0.042 & 5375 & 2.0 & 0.062 & 5250$^{\rm a}$ & 3.0 & 0.023 & 0.0  \\
G8 III & 5125 & 2.0 & 0.034 & 5125 & 2.0 & 0.058 & 5125 & 2.0 & 0.011 & 0.0  \\
K0 III & 4875 & 1.5 & 0.046 & 4875 & 1.5 & 0.075 & 4875 & 2.0 & 0.022 & 0.0  \\
K1 III & 4562.5 & 3.0 & 0.068 & 4562.5 & 3.0 & 0.114 & 4625 & 2.5 & 0.017 & 0.0  \\ 
K2 III & 4500 & 2.0 & 0.056 & 4500 & 2.0 & 0.094 & 4500 & 2.0 & 0.019 & 0.0  \\
K3-4 III & 4250 & 1.25 & 0.095 & 4250 & 1.25 & 0.154 & 4187.5 & 1.0 & 0.032 & 0.0  \\
\hline
G0 V   & 5937.5 & 4.0 & 0.035 & 5937.5 & 4.0 & 0.059 & 6000 & 5.0 & 0.013 & 0.0 \\
G4-5 V & 5625 & 4.0 & 0.052 & 5625 & 4.0 & 0.078 & 5750 & 5.0 & 0.028 & 0.0  \\
\hline
G8 III  & 4750 & 2.0 & 0.050 & 4750 & 2.0 & 0.077 & 4625 & 2.5 & 0.016 & -0.5  \\ 
K1.5 III\tablenotemark{b} & 4312.5 & 1.0 & 0.069 & 4187.5 & 2.0 & 0.117 & 4312.5 & 1.0 & 0.023 & -0.5  \\
\enddata
\tablenotetext{a}{Value suspect - see text.}
\tablenotetext{b}{Arcturus, $\alpha$ Boo}
\label{tabstats1}
\end{deluxetable} 

\clearpage

\begin{deluxetable}{lrrrrrrrrrrr}
\tablecolumns{11}
\tablecaption{NLTE models: Same as Table \ref{tabstats1}. }
\tablehead{
\colhead{}              & \multicolumn{3}{c}{Total SED}      &  \multicolumn{3}{c}{Blue}                     & \multicolumn{3}{c}{Red}                      & \colhead{} \\ 
\colhead{Spectral type} & \colhead{$T_{\rm eff}$} & \colhead{$\log g$} & \colhead{$\sigma_{\rm Min}$} & \colhead{$T_{\rm eff}$} & \colhead{$\log g$} & \colhead{$\sigma_{\rm Min}$} & \colhead{$T_{\rm eff}$} & \colhead{$\log g$} & \colhead{$\sigma_{\rm Min}$} & \colhead{$[{{\rm A}\over{\rm H}}]$} \\
} 
\startdata
G5 III & 5312.5$^{\rm a}$ & 2.0 & 0.042 & 5312.5 & 2.0 & 0.063 & 5250$^{\rm a}$ & 3.0 & 0.022 & 0.0 \\
G8 III & 5062.5 & 2.0 & 0.034 & 5062.5 & 2.0 & 0.058 & 5062.5 & 2.0 & 0.010 & 0.0\\
K0 III & 4812.5 & 1.5 & 0.044 & 4812.5 & 1.5 & 0.069 & 4812.5 & 2.0 & 0.021 & 0.0\\
K1 III & 4562.5 & 2.5 & 0.069 & 4500 & 3.0 & 0.114 & 4562.5 & 2.0 & 0.016 & 0.0\\
K2 III & 4500 & 2.0 & 0.057 & 4500 & 2.0 & 0.095 & 4437.5 & 1.5 & 0.020  & 0.0\\
K3-4 III & 4187.5 & 1.5 & 0.094 & 4187.5 & 1.5 & 0.154 & 4125 & 1.0 & 0.031 & 0.0\\
\hline
G0 V   & 5937.5 & 4.0 & 0.036 & 5937.5 & 4.0 & 0.060 & 5937.5 & 4.5 & 0.012  & 0.0\\
G4-5 V & 5625 & 4.0 & 0.053 & 5562.5 & 4.0 & 0.081 & 5687.5 & 5.0 & 0.028 & 0.0\\ 
\hline
G8 III  &                 4687.5 & 2.0 & 0.050 & 4687.5 & 2.0 & 0.078 & 4562.5 & 2.5 & 0.015 & -0.5\\ 
K1.5 III\tablenotemark{b} &  4187.5 & 1.5 & 0.073 & 4187.5 & 1.5 & 0.120 & 4250 & 1.25 & 0.023 & -0.5\\
\enddata
\tablenotetext{a}{Value suspect - see text.}
\tablenotetext{b}{Arcturus, $\alpha$ Boo}
\label{tabstats2}
\end{deluxetable} 

\clearpage
%
\begin{deluxetable}{lrrrrrrr}
\tablecolumns{8}
\tablecaption{Comparison with less model-dependent $T_{\rm eff}$ calibrations of RM05 and B10.}
\tablehead{
\colhead{ }             & \colhead{ }   & \multicolumn{2}{c}{LTE}     & \multicolumn{2}{c}{NLTE}     & \colhead{ }    & \colhead{ } \\ 
\colhead{Spectral type} & \colhead{B-V} ($\sigma$) & \colhead{Blue } & \colhead{Red } & \colhead{Blue } & \colhead{Red } & \colhead{RM05} & \colhead{B10} } 
\startdata
G5 III & 0.882 (0.019) & 5375 & \nodata & 5312.5 & \nodata & 5137 &  \nodata \\
G8 III & 0.930 (0.004) & 5125 & 5125 & 5062.5 & 5062.5 & 4964 &  \nodata  \\   
K0 III & 1.043 (0.002) & 4875 & 4875 & 4812.5 & 4812.5 & 4721 &  \nodata\\  
K1 III & 1.115 (0.003) & 4562.5 & 4625 & 4500 & 4562.5 & 4592 &  4737 \\     
K2 III & 1.160 (0.006) & 4500 & 4500 & 4500 & 4437.5 & 4531 &  4562 \\   
K3-4 III & 1.408 (0.014) & 4250 & 4187.5 & 4187.5 & 4125 & 4118 &  4134 \\   
\hline
G0 V & 0.595 (0.004) & 5937.5 & 6000 & 5937.5 & 5937.5 & 5864 &  \nodata \\   
G4-5 V & 0.695 (0.010) & 5625 & 5750 & 5562.5 & 5687.5 & 5519 &  \nodata\\
\hline
G8 III-0.5 & 1.010 (0.000) & 4750 & 4625 & 4687.5 & 4562.5 & 4684 & \nodata\\  
K1.5 III-0.5$^{\rm a}$ & 1.211 (0.009) & 4187.5 & 4312.5 & 4187.5 & 4250 & 4332 & 4386\\     
\enddata
\tablenotetext{a}{Arcturus, $\alpha$ Boo}
\label{tabcomp1}
\end{deluxetable}

\end{document}